\documentclass[onecolumn]{aa}
\usepackage{graphicx}
\usepackage{times}
\usepackage{aalongtable}
\usepackage{natbib}
\usepackage{psfig}

\bibpunct{(}{)}{;}{a}{}{,} 

\def\5{\footnotesize V\normalsize}
\def\4{\footnotesize IV\normalsize}
\def\3{\footnotesize III\normalsize}
\def\2{\footnotesize II\normalsize}
\def\1{\footnotesize I\normalsize}
\def\lam{$\lambda$}
\def\kms{$\mbox{km s}^{-1}$}
\def\p{$\phantom{:}$}

\def\pp{$\phantom{-}$}
\def\o{$\phantom{1}$}

\begin{document}

\title{The VLT-FLAMES Survey of Massive Stars: Observations in the Galactic 
Clusters NGC\,3293, NGC\,4755 and NGC\,6611\footnote{Based on observations 
at the European Southern Observatory in programmes 171.D-0237 \& 073.D-0234.}}

\author{C.~J.~Evans\inst{1}, 
S.~J.~Smartt\inst{2}, 
J.-K.~Lee\inst{2}, 
D.~J.~Lennon\inst{1}, 
A.~Kaufer\inst{3}, 
P.~L.~Dufton\inst{2},
C.~Trundle\inst{4}, 
A.~Herrero\inst{4,5}, 
S.~Sim${\rm \acute{o}}$n-D\'{\i}az\inst{4}, 
A.~de~Koter\inst{6}, 
W.-R.~Hamann\inst{7},  
M.~A.~Hendry\inst{8}, 
I.~Hunter\inst{2},
M.~J.~Irwin\inst{8}, 
A.~J.~Korn\inst{9}, 
R.-P.~Kudritzki\inst{10},
N.~Langer\inst{11}, 
M.~R.~Mokiem\inst{6}, 
F.~Najarro\inst{12}, 
A.~W.~A.~Pauldrach\inst{13}, 
N.~Przybilla\inst{14}, 
J.~Puls\inst{13}, 
R.~S.~I.~Ryans\inst{2}, 
M.~A.~Urbaneja\inst{10}, 
K.~A.~Venn\inst{15}, 
M.~R.~Villamariz\inst{4}
}

\offprints{C.~J.~Evans at cje@ing.iac.es}

\authorrunning{C.~J.~Evans et al.}
   
\titlerunning{FLAMES Survey of Massive Stars: MW Clusters}
   
\institute{The Isaac Newton Group of Telescopes,
           Apartado de Correos 321, E-38700,
           Santa Cruz de La Palma, Canary Islands, Spain
             \and 
           The Department of Pure and Applied Physics,
           The Queen's University of Belfast,
           Belfast BT7 1NN, Northern Ireland, UK
             \and
           European Southern Observatory, Alonso de Cordova 3107, Santiago 19, Chile
             \and
           Instituto de Astrof\'{\i}sica de Canarias, 
           E-38200, La Laguna, Tenerife, Spain
             \and
           Dept. de Astrof\'{\i}sica, Universidad de La Laguna,
           Avda. Astrof\'{\i}sico Francisco S$\acute{\rm a}$nchez, 
           s/n, E-38071 La Laguna, Spain
             \and
           Astronomical Institute Anton Pannekoek, University of Amsterdam, 
           Kruislaan 403, 1098 SJ Amsterdam, The Netherlands
             \and
           Lehrstuhl Astrophysik der Universit$\ddot{\rm a}$t Potsdam, Am Neuen Palais 10, 
           D-14469 Potsdam, Germany
             \and
           Institute of Astronomy, University of Cambridge,
           Madingley Road, Cambridge, CB3 0HA, UK
             \and
           Uppsala Astronomical Observatory, Box 515, SE-75120, Uppsala, Sweden
             \and
           Institute for Astronomy, 2680 Woodlawn Drive, 
           Honolulu, HI 96822, USA
             \and
           Astronomical Institute, Utrecht University, Princetonplein 5,
           NL-3584 CC Utrecht, The Netherlands
             \and
           Instituto de Estructura de la Materia, CSIC,
           C/ Serrano 121, E-28006 Madrid, Spain
             \and
           Universit$\ddot{\rm a}$ts-Sternwarte M$\ddot{\rm u}$nchen, 
           Scheinerstr. 1, D-81679, M$\ddot{\rm u}$nchen, Germany
             \and
           Dr. Remeis-Sternwarte Bamberg, Sternwartstr. 7, D-96049 Bamberg, Germany 
             \and
           Macalester College, 1600 Grand Avenue, Saint Paul, MN 55105, USA 
}
\date{}

\abstract{
We introduce a new survey of massive stars in the Galaxy and the
Magellanic Clouds using the Fibre Large Array Multi-Element
Spectrograph (FLAMES) instrument at the Very Large Telescope (VLT).
Here we present observations of 269 Galactic stars with the
FLAMES-Giraffe Spectrograph ($R \simeq 25,000$), in fields centered on
the open clusters NGC\,3293, NGC\,4755 and NGC\,6611.  These data are
supplemented by a further 50 targets observed with the Fibre-Fed
Extended Range Optical Spectrograph (FEROS, $R$ = 48,000).  Following
a description of our scientific motivations and target selection
criteria, the data reduction methods are described; of critical importance
the FLAMES reduction pipeline is found to yield spectra that are in
excellent agreement with less automated methods.  Spectral
classifications and radial velocity measurements are presented for
each star, with particular attention paid to morphological
peculiarities and evidence of binarity.  These observations represent
a significant increase in the known spectral content of NGC\,3293
and NGC\,4755, and will serve as standards against which our subsequent
FLAMES observations in the Magellanic Clouds will be compared.}

\maketitle 

\section{Introduction}
The Fibre Large Array Multi-Element Spectrograph (FLAMES) at the Very
Large Telescope (VLT) provides multi-object spectroscopy over a
corrected 25$'$ diameter field-of-view.  As part of a European
Southern Observatory (ESO) Large Programme, comprising over 100 hours
of VLT time, we have observed in excess of 50 O-type stars and 500
B-type stars, in a total of seven clusters distributed over the Galaxy
and the Magellanic Clouds.  This unprecedented survey of massive
stars, and its homogeneous spectroscopic analysis using
state-of-the-art model atmospheres, will be the subject of a series of
papers.  This first paper gives an overview of the scientific goals of
the project, discusses the data reduction techniques employed for the
survey, and presents the data for the three Galactic clusters.

A diverse range of topics depend critically on our understanding of
the evolution of massive stars.  Through their stellar winds and as
supernova explosions they are the main source of kinetic energy into
the interstellar medium of their host galaxies, and they are
responsible for the continuous build-up of elements heavier than
helium (`metals') from successive generations of star formation
\citep[e.g.][]{shap04}.  Indeed, one of the key challenges is to 
fully understand the role of metallicity on stellar evolution, to
place better constraints on extremely low-metallicity stellar models
\citep[e.g.][]{kud02}.  Theoretical simulations of star formation in
the early Universe have suggested that masses larger than
100~M$_{\odot}$ are strongly favoured \citep{bcl02,abn02}, and
these `zero-metallicity' (or Population III) stars are considered to
be prime candidates for the re-ionization of the Universe
\citep[e.g.][]{hmkh01,wl03} that may have happened at
redshifts greater than six \citep{bfw01}.  

Furthermore, with ongoing studies of the nature of supernova progenitors
\citep[e.g.][]{smh04}, and following recent discoveries that some of
the most spectacular supernovae appear to be related to $\gamma$-ray bursts
\citep[e.g.][]{gal98,bk02,hjorth}, understanding the evolution of
massive stars will gain even greater importance in the near future.
However, fundamental problems remain as to how the evolution of
massive stars depends on the physics of mass-loss and rotation, and
the role of metallicity.  The FLAMES survey of massive stars aims at a
comprehensive study of these processes. Some of the main scientific
issues that we will address with the project are:
\begin{itemize}
\item{Stellar rotation and chemical abundances:
The inclusion of the effects of stellar rotation in evolutionary
models leads to the important prediction of enhanced surface 
abundances of helium and nitrogen \citep[e.g.][]{hl00,mm00}, with the
enhancements predicted to be more significant at lower metallicity
\citep{mm01}.  Results from model atmosphere analyses of O- and B-type 
supergiants in the Magellanic Clouds \citep{paul,jc03,jdh03,ecfh,tl04,wal04,tl05} find
evidence of significant nitrogen enrichment, that could in principle be attributed
to mixing occurring in stars with initially high rotational velocities.
However, none of the stars analysed to date have particularly large
projected rotational velocities and it seems likely that either they
have `spun down' more quickly than the models predict, or that
rotationally-induced mixing is more effective than expected, even at relatively
moderate velocities and at earlier evolutionary phases.  The FLAMES
survey will be used to determine stellar rotational velocities and to then
provide a detailed study of their correlation with helium and CNO abundances.}
\\
\item{The dependence of stellar wind mass-loss rates on metallicity:
Accurate interpretation of the integrated spectra of starbursts 
\citep[e.g., ][]{vlh04} and high-redshift galaxies 
\citep[e.g., ][]{max, rix} requires detailed knowledge of the dependence 
of stellar mass-loss rates, $\dot{M}$ with metallicity.  Radiatively
driven wind theory predicts that $\dot{M}$ should scale as
$Z^{0.5-0.7}$ \citep[e.g., ][]{kpp,v01}.  Observational evidence of
this is still rather qualitative \citep[e.g.][]{elw04} and recent
results in the Magellanic Clouds \citep[e.g., ][]{ecfh, tl04} have
stressed the need for a large, well-sampled, homongenous study.  The
FLAMES sample will yield mass-loss rates (and chemical abundances) for
a significant number of early-type stars in different environments,
and will be well-suited to test the metallicity dependence; this is
essential if we are to correctly include its effects in stellar
evolution models and, ultimately, in population synthesis codes.  }
\\
\item{Calibration of the wind-momentum luminosity relationship (WLR):
The dependence of the modified wind momentum on stellar luminosity for
OBA-type stars \citep[e.g., ][]{klp95, p96, k99} potentially offers a
purely spectroscopic method of distance determination in the Local
Group and beyond.  Recent analyses by \citet{mp04} and \citet{rp04}
have revisited this relationship for Galactic O-type stars; the FLAMES
survey will enable a more precise calibration of this relationship in
the Clouds.
}
\end{itemize}
The clusters observed with FLAMES are summarized in Table
\ref{fields}.  The earliest (main sequence) spectral types observed in
each cluster are included to give an indication of their relative
ages.  A typical FLAMES field yields 110-120 targets and at each
metallicity we have targeted a young cluster and an older cluster (two
in the case of the Milky Way).

In the current paper we discuss the methods used for target selection
and detail the data reduction techniques employed for the survey.  We
also present spectral classifications, radial velocities and
cross-identifications with previous catalogues for our targets in the
three Galactic clusters: NGC\,3293, NGC\,4755 ({\it aka} The Jewel
Box), and NGC\,6611 ({\it aka} The Eagle Nebula; Messier 16).  The
data presented here also include observations of the brightest cluster
members with the Fibre-Fed, Extended Range Optical Spectrograph
(FEROS) at the 2.2-m Max Planck Gesellschaft (MPG)/ESO telescope.  A
thorough analysis of the rotational velocities of the Galactic sample
will be given elsewhere (Lee et al., in preparation) and similar
catalogues for the Magellanic Cloud targets will be presented in a
further publication (Evans et al., in preparation).

\begin{table}
\caption[]{Galactic and Magellanic Cloud clusters observed with FLAMES.
The earliest main-sequence object observed in each cluster is given as
an indication of their relative ages.}
\label{fields}
\begin{center}
\begin{tabular}{lll}
\hline\hline
Cluster &   Galaxy    & Earliest main-sequence type \\
\hline
NGC3293 &   Milky Way & B1 V \citep{f3293}\\
NGC4755 &   Milky Way & B0.5 V \citep{f4755}\\
NGC6611 &   Milky Way & O5 V \citep{hmsm}\\
NGC330  &   SMC       & B0 V \citep{l93}\\
NGC346  &   SMC       & O4 V \citep{wal00}\\
NGC2004 &   LMC       & early B\\
LH9/10  &   LMC       & O3 V \citep{p92}\\
\hline
\end{tabular}
\end{center}
\end{table}

\section{Target Selection}
\label{targets}

\subsection{WFI Photometry}
Astrometry and photometry for our target fields were acquired from
observations for the ESO Imaging Survey (EIS) that used the Wide Field
Imager (WFI) at the 2.2-m MPG/ESO telescope \citep{eis}.  The relevant
$B$ ($B$/99, ESO\#842) and $V$ ($V$/89, ESO\#843) science and
calibration images were taken from the ESO archive and reduced using a
modified version of the Isaac Newton Telescope--Wide Field Camera
(INT--WFC) data reduction pipeline \citep{mji_wfc}.  Multiple
exposures of the fields (aimed at shallow and deeper magnitudes) were
combined to produce the final photometry; the observations are
summarised in Table \ref{obs_wfi}.  

\begin{table}
\caption[]{Summary of the relevant EIS pre-FLAMES observations of Galactic fields
used for target selection and astrometry.}
\label{obs_wfi}
\begin{center}
\begin{tabular}{lccccll}
\hline\hline
Cluster   &  EIS field & \multicolumn{2}{c}{Field Centre (J2000)} &  Date          & \multicolumn{2}{c}{Exposure time (sec)}\\
          &            & $\alpha$    & $\delta$                   &                &   $B$      &  $V$      \\
\hline
NGC\,3293 &  OC15      & 10 35 28.0  & $-$58 21 14                &  2000-02-26    &  10+30+240+240 & 10+30+240+240 \\
NGC\,4755 &  OC22      & 12 53 13.0  & $-$60 28 06                &  2000-02-26    &  30+240+240    & 30+240+240    \\  
NGC\,6611 &  OC32      & 18 18 51.9  & $-$13 46 36                &  2000-07-29    &  30+240+240    & 30+240+240    \\
\hline
\end{tabular}
\end{center}
\end{table}

Photometric transformations for the WFI data to the Johnson-Cousins
system were determined for each cluster from published photometry,
using visually matched stars -- in the absence of observations of
standard fields this `bootstrap' approach was both necessary and
sufficient for our requirements.  The brightest targets in each
cluster were saturated, even in the short WFI exposures,
although the astrometry of these sources was more than adequate for
positioning of the FLAMES fibres.  The photometric transformations were
determined using those targets that displayed a linear response i.e.,
$V >$ 10.75$^{\rm m}$ in NGC\,3293 and $V >$ 11.5$^{\rm m}$ in
NGC\,4755 and NGC\,6611.

\begin{figure*}
\begin{center}
\includegraphics{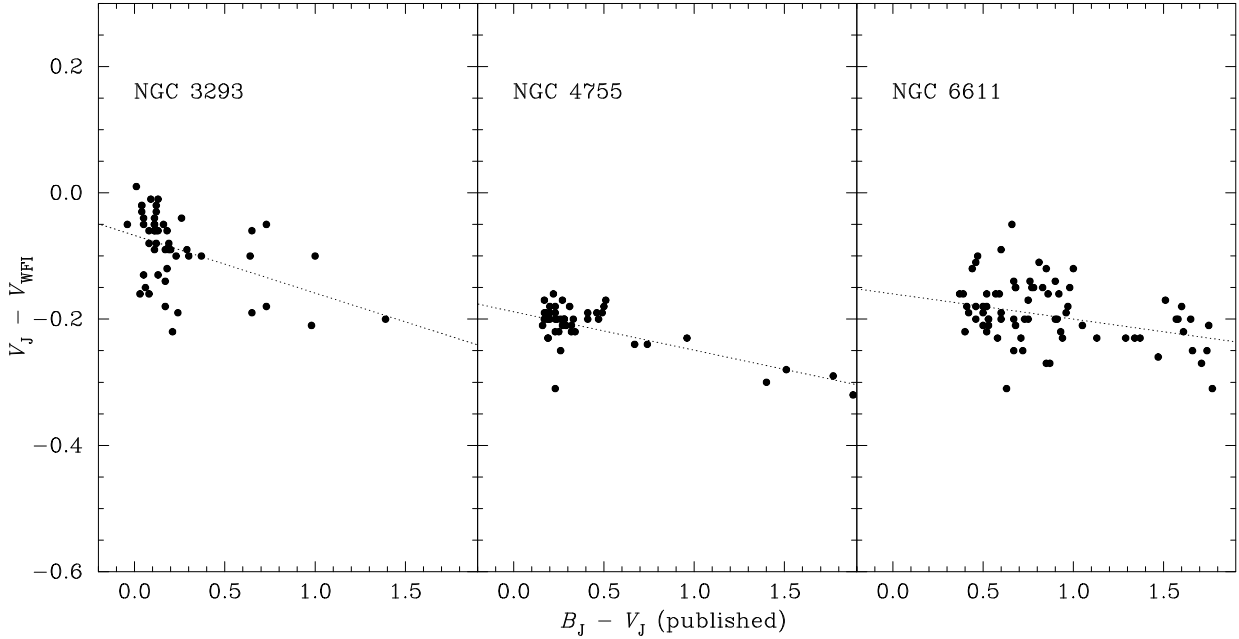}
\caption{Comparison of $V_{\rm J} - V_{\rm WFI}$ with published colours from 
\citet[][ NGC\,3293]{turn80}, \citet[][ NGC\,4755]{sb01} and 
\citet[][ NGC\,6611]{hmsm}.  In addition to the zero-point offset, there is a
relatively weak colour term between the WFI data and published
results.}
\label{v}
\end{center}
\end{figure*}

\begin{figure*}
\begin{center}
\includegraphics{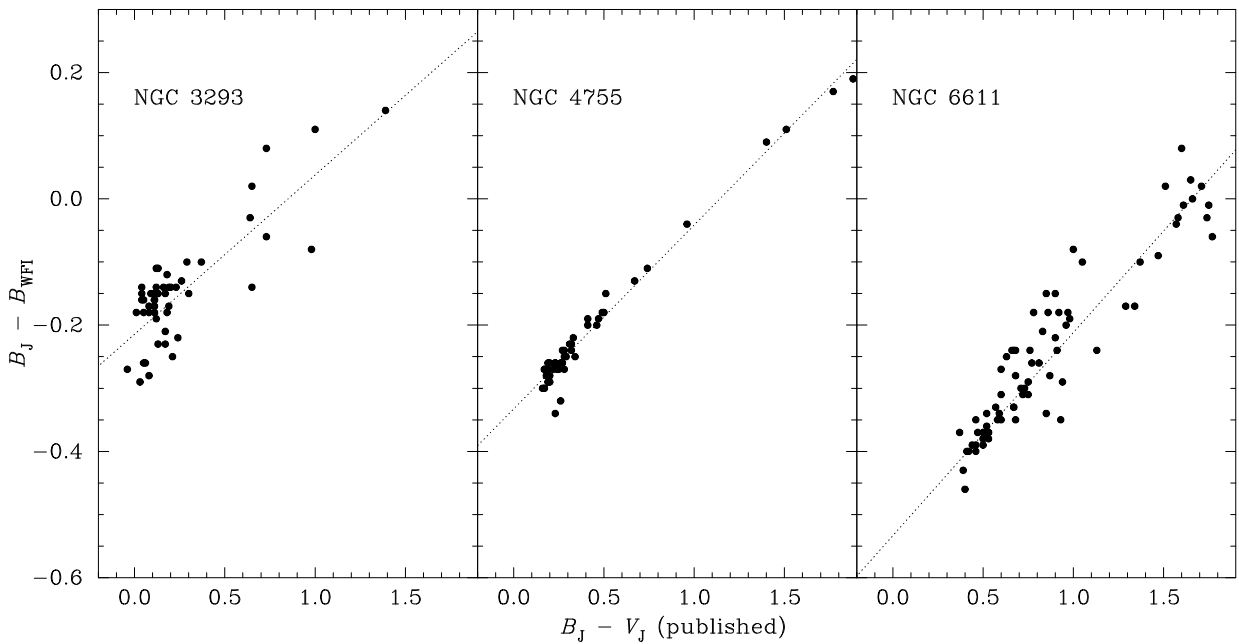}
\caption{Comparison of $B_{\rm J} - B_{\rm WFI}$ with published colours from 
\citet[][ NGC\,3293]{turn80}, \citet[][ NGC\,4755]{sb01} and 
\citet[][ NGC\,6611]{hmsm}.  In addition to the zero-point offset, there is a
colour term between the WFI data and published results.}
\label{b}
\end{center}
\end{figure*}

In Figures \ref{v} and \ref{b} we show fits to the colour terms in $V$
and $B$; published photometry was taken from 
\citet[][ NGC\,3293]{turn80}, \citet[][ NGC\,4755]{sb01}, and 
\citet[][ NGC\,6611]{hmsm}.  The transformation equations found (from 51 stars) for 
NGC\,3293 were:
\begin{equation}
V_{\rm J} = V_{\rm WFI} - 0.09\times(B - V)_{\rm J} - 0.07, 
\end{equation}
\begin{equation}
B_{\rm J} = B_{\rm WFI} + 0.25\times(B - V)_{\rm J} - 0.21.
\end{equation}
Similarly, the solutions (from 48 stars) for NGC\,4755 were:
\begin{equation}
V_{\rm J} = V_{\rm WFI} - 0.06\times(B - V)_{\rm J} - 0.19, 
\end{equation}
\begin{equation}
B_{\rm J} = B_{\rm WFI} + 0.29\times(B - V)_{\rm J} - 0.33.
\end{equation}
Lastly, the solutions (from 75 stars) for NGC\,6611 were:
\begin{equation}
V_{\rm J} = V_{\rm WFI} - 0.04\times(B - V)_{\rm J} - 0.16, 
\end{equation}
\begin{equation}
B_{\rm J} = B_{\rm WFI} + 0.32\times(B - V)_{\rm J} - 0.53.
\end{equation}
These colour terms are comparable to those found by \citet{eis} for
the pre-FLAMES WFI survey.  In comparison with previously
published values, after transformation we find mean (absolute)
differences of $\sim 0.03^{\rm m}$, with $\sigma \sim 0.03^{\rm m}$
for both $V$ and $(B - V)$ in NGC\,3293 and NGC\,4755; in NGC\,6611 we
find differences of $0.04^{\rm m}$, with $\sigma = 0.04^{\rm m}$ in
both $V$ and $(B - V)$.  Such accuracy is adequate for our purposes of
target selection.  We note that the fit for NGC\,3293 in Figure \ref{v} is not 
biassed by the one star at $(B - V) >$~1; the same colour term is found regardless
of its inclusion.

Following transformation to the standard system, the observed stars in
each cluster were assigned running numbers on the basis of the
$V$-band magnitudes, with `3293-001' the brightest star observed in
NGC\,3293 and so on.  This is not strictly the case for the stars in
NGC\,6611, following refinement of the calibrations (using an enlarged
sample of published values) after the assignment of the running
numbers.

In the process of determining the photometric transformations we noticed that
two stars in NGC\,4755, namely IV-11 and IV-17 in the numbering system of
\citet[][]{as58}, are listed by \citet{sb01} as 
having $V \sim$ 14$^{\rm m}$.  These are significantly discrepant from
the (albeit photographic) values given by \citeauthor{as58} ($V
=$~11.35 and 9.76$^{\rm m}$ respectively).  Photographic (IV-11: $V
=$~11.41$^{\rm m}$) and photoelectric (IV-17: $V =$~9.96$^{\rm m}$)
results from \citet{dk84} roughly tally with the \citeauthor{as58} values,
and the WFI observations confirm that they are much brighter than 14th
magnitude.  We suggest that these may be misidentifications by
\citet{sb01} rather than significant, intrinsic variability.

More curious is the case of Walker 442 in NGC\,6611, with \citet{the90}
reporting $V =$~8.26. The photographic observations by
\citet[][ $V =$~15.09$^{\rm m}$]{w61} and the value from the 
WFI data ($V =$~14.99$^{\rm m}$) are both much fainter for this star.
Though the star is not explicitly cross-referenced, \citet{hmsm} also
give $V =$~15.00 for a star at $\alpha =$~18$^{\rm h}$~19$^{\rm
m}$~0.86$^{\rm s}$, $\delta =$~$-$13$^{\circ}$~48$^{'}$~47.2$^{''}$
(J2000), almost certainly W442.  It is hard to reconcile these
observations with the \citet{the90} value which, if real, would make
W442 one of the visually brightest stars in the cluster.  The sources
of this discrepancy are not clear, but we note that the SIMBAD
database reports the \citeauthor{the90} results.

\subsection{FLAMES-Giraffe spectroscopy}

The primary element of FLAMES is the Giraffe spectrograph.  Our
programme employed the mode of Giraffe in which it is fed by 132
so-called `Medusa' fibres.  Input catalogues of possible targets were
compiled for each cluster from the WFI photometry and astrometry;
these were then used to allocate the Medusa fibres with the FLAMES
Fibre Positioner Observation Support Software (FPOSS).  The philosophy
behind compilation of Giraffe targets was to select bright blue stars,
with cuts of $(B - V)_{0} <$ 0.00$^{\rm m}$ (corresponding to the threshold
between B and A spectral types) and to a faint limit such that the
main sequence was sampled down to $\sim$B5 (corresponding to M$_V \sim
-1$).

An important consideration for application of these selection criteria
in the Galactic clusters is an estimate of the extinction and
reddening.  Estimates of $E(B - V)$ were incorporated into our
selection routines, adopting $E(B - V) =$ 0.30$^{\rm m}$ for NGC\,3293
\citep[e.g.,][]{turn80} and 0.40$^{\rm m}$ for NGC\,4755 \citep[e.g.,][]{dewaard}.  
The treatment of reddening in NGC\,6611 was somewhat more complicated;
the extinction is large and strongly variable
across the cluster and a value of $E(B - V) =$ 1.20$^{\rm m}$ was
adopted, which is at the upper end of those found by \citet{hmsm}.
This value ensured inclusion of the majority of the early-type objects
(necessary for a reasonable density of targets across the FLAMES
field-of-view), but contamination by (less-reddened) later-type stars
can then become a problem.  Indeed, as demonstrated by the spectral
types present in our final sample (see Table \ref{6611}), selection of
solely early-type objects in such a cluster is extremely difficult
without previous spectroscopic information.

Following compilation of the input catalogues, FPOSS was used to
assign the Medusa fibres.  In addition to selection effects such as
avoiding fibre collisions and the physical size of the fibre buttons,
the final spectral samples are influenced by some additional external
constraints.  Where published spectral types were available (and the
object was of interest) a higher weight was assigned to the target
when using FPOSS.  Similarly, known Be-type stars were (in general)
deliberately omitted from the input catalogues used with FPOSS; this
is primarily relevant to configuration of our Magellanic Cloud fields
but is mentioned here to note that this survey is $not$ suitable for
statistical studies of the incidence of Be-type stars.

The brightest stars in each cluster were deliberately omitted from the
FLAMES observations as they were too bright for the dynamic range of
the instrument; such stars would simply saturate during the exposure
times needed for the fainter B-type main-sequence stars.  These stars
were observed separately using FEROS at the 2.2-m MPG/ESO telescope
(see Section \ref{feros}).

\begin{figure}
\begin{center}
\includegraphics{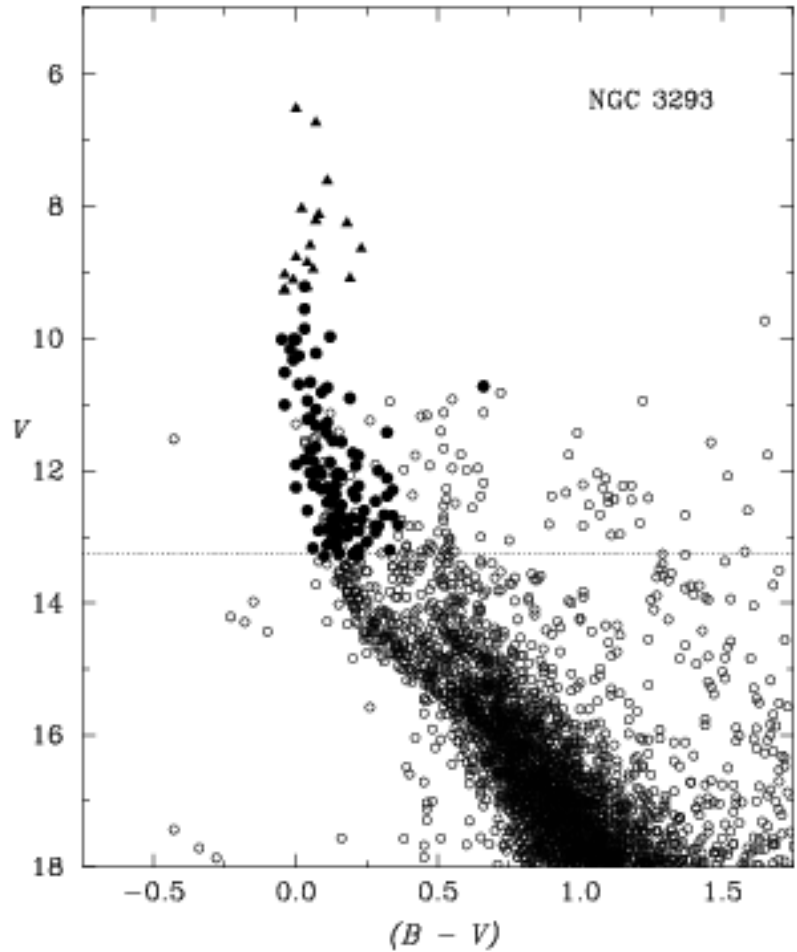}
\caption{Colour-magnitude diagram for all objects in the FLAMES
field-of-view for NGC\,3293.  The marked targets were observed
with FLAMES-Giraffe (solid circles) or with FEROS (solid
triangles).
\label{3293_bv}}
\end{center}
\end{figure}

\begin{figure}
\begin{center}
\includegraphics{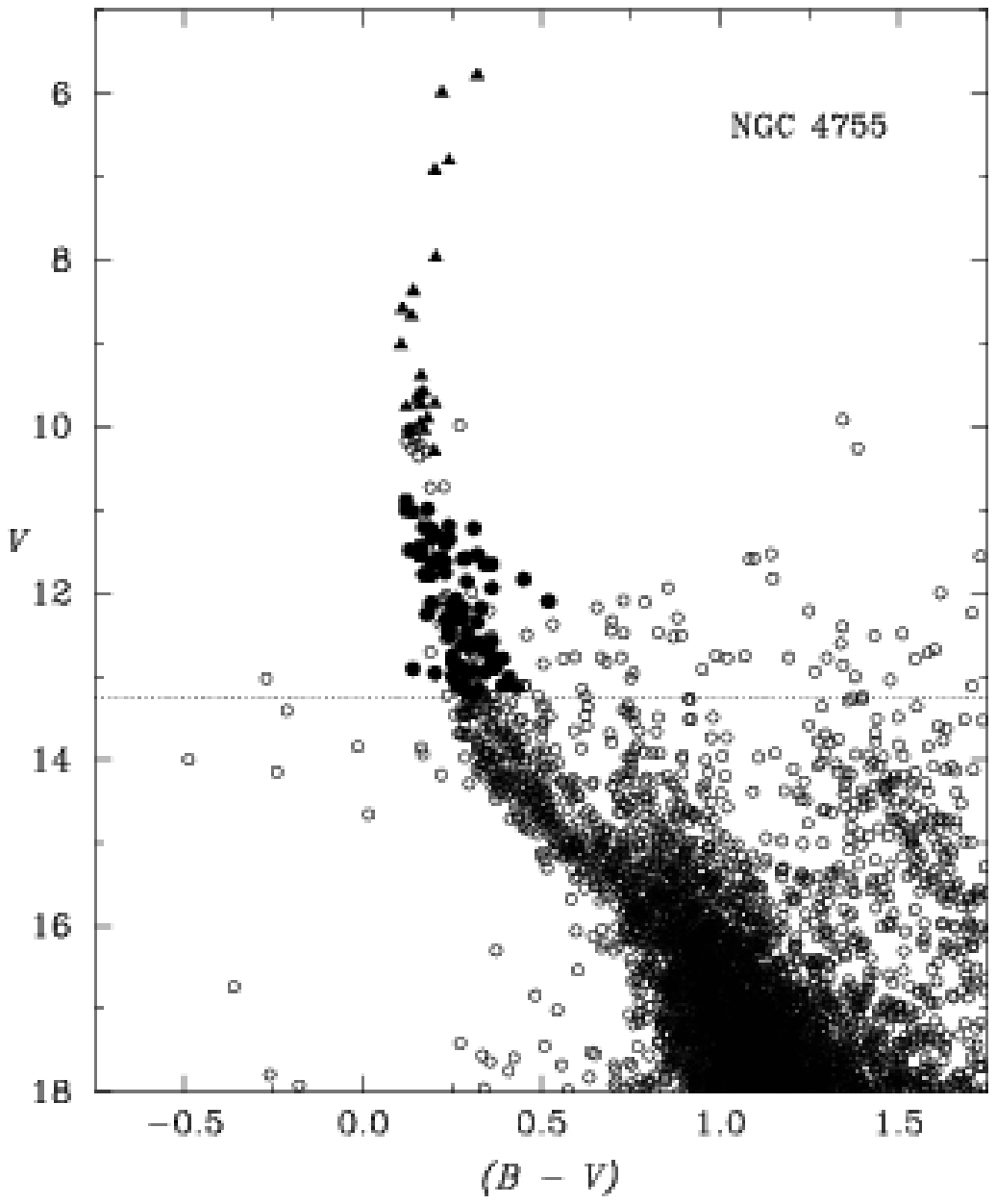}
\caption{Colour-magnitude diagram for all objects in the FLAMES
field-of-view for NGC\,4755.  The marked targets were observed
with FLAMES-Giraffe (solid circles) or with FEROS (solid
triangles).
\label{4755_bv}}
\end{center}
\end{figure}

\begin{figure}
\begin{center}
\includegraphics{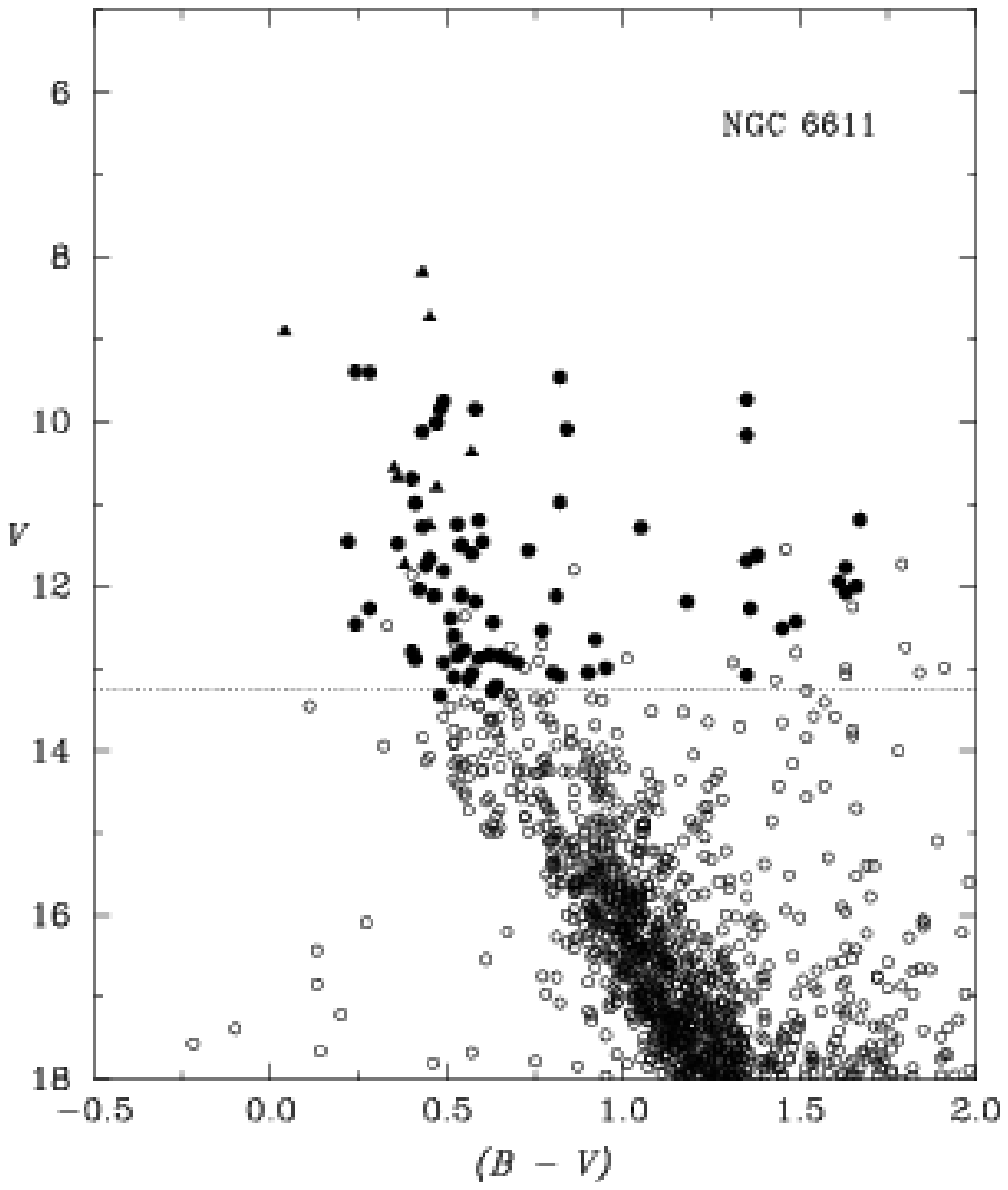}
\caption{Colour-magnitude diagram for all objects in the FLAMES
field-of-view for NGC\,6611.  The marked targets were observed
with FLAMES-Giraffe (solid circles) or with FEROS (solid
triangles).
\label{6611_bv}}
\end{center}
\end{figure}

The colour-magnitude diagram for all the stars in the FLAMES
field-of-view centered on NGC\,3293 is shown in Figure \ref{3293_bv}.
Magnitudes and colours for stars fainter than $V =$ 10.75$^{\rm m}$ are
from our WFI photometry (corrected to the Johnson system); with
photometry for the brighter stars from \citet{turn80} and
\citet{fm90}.  The faint limit of our targets ($V \sim$13.25$^{\rm
m}$) is marked by the dotted line.  The colour-magnitude diagrams for
NGC\,4755 and 6611 are shown in Figures \ref{4755_bv} and
\ref{6611_bv}, with the photometry for the brightest stars from
\citet{sb01} and \citet{hmsm} respectively.

\subsection{FEROS spectroscopy}
\label{feros}
The brightest targets in NGC\,3293 and NGC\,4755 were observed using
FEROS at the ESO/MPG 2.2-m telescope on 2004 July 4-6.  We also sought
observations of remaining targets of interest in the NGC\,6611 field
and, to increase the number of Galactic O-type stars in the survey, we
included three stars classified as O-type by \citet{hmsm} that were
not in the FLAMES field-of-view (6611-001, 6611-005, and 6611-045;
W412, HD\,168504, and W584 respectively).

FEROS is a fixed configuration instrument (with $R$ = 48,000), giving
a wide wavelength coverage of 3600-9200\,\AA\/ in one exposure.  The
FEROS data were reduced using the reduction pipeline that runs under
the MIDAS environment \citep{ak99}.  In the majority of cases the
signal-to-noise ratio of the FEROS data is in excess of 100 per
resolution element.  Targets observed with FEROS are marked in Figures
\ref{3293_bv}, \ref{4755_bv}, and \ref{6611_bv} as solid triangles.

\section{FLAMES-Giraffe Observations and Data Reduction}
\label{obs}
Each of the three Milky Way clusters was observed with one FLAMES
fibre configuration, centered on the cluster core.  Six of the
standard high-resolution Giraffe settings were used: HR02 (with a
central wavelength of \lam3958\,\AA), HR03 (\lam4124), HR04
(\lam4297), HR05 (\lam4471), HR06 (\lam4656), and HR14 (\lam6515).
The spectral range for each setting is approximately 200\,\AA, ie:
there is sufficient overlap between settings to give a continuous
spectrum of 3850-4755\,\AA, with additional coverage of
6380-6620\,\AA, primarily to observe the H$\alpha$ Balmer line.  The
observations are summarized in Table \ref{obs_flames}, with details of
the individual Galactic targets given in Tables \ref{3293},
\ref{4755} and \ref{6611}.

\begin{table}
\caption[]{Summary of VLT-FLAMES observations of Galactic clusters.  The
exposure times are for {\it each} wavelength setting.}
\label{obs_flames}
\begin{center}
\begin{tabular}{lcclcl}
\hline\hline
Cluster   &  \multicolumn{2}{c}{Field Centre (J2000.0)} & Date       & Exp. time    & Giraffe settings    \\
          &  $\alpha$        & $\delta$                 &            & (sec)        &                     \\
\hline	    
NGC\,3293 &  10 35 48.7      & $-$58 13 34              & 2003-04-14 & 795          & HR02/03/04/05/06/14 \\
NGC\,4755 &  12 53 42.0      & $-$60 22 15              & 2003-04-13 & 795          & HR02/03/04/05/06/14 \\
NGC\,6611 &  18 18 50.0      & $-$13 46 44              & 2003-07-15 & 2$\times$375 & HR02/03/04/05/06/14 \\
NGC\,6611 &  18 18 50.0      & $-$13 46 44              & 2003-07-18 & 2$\times$375 & HR05/06/14 \\
\hline
\end{tabular}
\end{center}
\end{table}

The Galactic FLAMES data were reduced using IRAF \footnote{{\sc iraf} is
distributed by the National Optical Astronomy Observatories, which are
operated by the Association of Universities for Research in Astronomy,
Inc., under cooperative agreement with the National Science
Foundation.} (v2.11).  Standard routines were used for bias
subtraction, fibre extraction and wavelength calibration.  To remove
the flat-field response from the science data, each extracted spectrum
was divided through by the normalised flat-field spectrum (taken as
part of the daytime calibrations) from the same fibre.  Further
manipulation of the data was undertaken using the {\sc STARLINK}
package {\sc DIPSO}.

The Galactic targets are relatively bright and sky subtraction was not
a critical issue, however such methods were fully explored for these
data as it will be more important for the fields observed in the Magellanic
Clouds.  A master sky spectrum was created from merging the available
sky spectra, rejecting those contaminated with a significantly larger
than average Balmer line emission component or those contaminated by
arc lines (see Section~\ref{arc_probs}).  In a cluster such as
NGC\,6611 there is considerable variation in the nebular emission
across very small scales (let alone across the FLAMES field-of-view)
and the subtraction of such features will not be 100$\%$ accurate; an
inherent drawback of multi-fibre spectrscopy.  The logic behind
rejecting those sky spectra with significantly large nebular emission
is driven by our desire to minimise over-subtraction in the line
cores, although this proved unavoidable in some of the NGC\,6611
targets.  The extracted flat-field fibres were used to determine
relative throughputs (compared to the median number of counts) for
each fibre; the master sky spectrum was then appropriately scaled
prior to subtraction from the science spectrum.

The exact resolving power of the grating varies with the different
central wavelength settings.  The mean FWHM of the arc-lines in the
central science fibre at each setting (averaged over the three
observed fields) is summarized in Table~\ref{arc_fwhm}.  The typical
signal-to-noise of the data ranges from 100-150 per resolution
element.

\begin{table}
\caption[]{Summary of the mean FWHM of the arc lines and effective
resolving power, $R$, at each Giraffe central wavelength setting, \lam$_c$.}
\label{arc_fwhm}
\begin{center}
\begin{tabular}{lcccc}
\hline\hline
Setting & \lam$_c$ & \multicolumn{2}{c}{FWHM}& $R$\\
        & (\AA)    & (\AA) & (pixels) & \\
\hline 
HR02    & 3958     & 0.19  & 3.7    & 20,850\\
HR03    & 4124     & 0.15  & 3.6    & 27,500\\
HR04    & 4297     & 0.19  & 3.6    & 22,600\\
HR05    & 4471     & 0.16  & 3.5    & 27,950\\
HR06    & 4656     & 0.20  & 3.6    & 23,300\\
HR14    & 6515     & 0.22  & 3.6    & 29,600\\
\hline
\end{tabular}
\end{center}
\end{table}

For classification purposes the spectra for each region were rectified
automatically (using pre-defined continuum regions) and then merged
(using {\sc DIPSO} routines developed by I. D. Howarth for echelle/fibre
reductions) to give a continuous spectrum of 3850-4755\,\AA, with
additional coverage of 6380-6620\,\AA.

\subsection{Contamination by simultaneous arc-line fibres}
\label{arc_probs}
For programmes that demand high-accuracy radial velocity measurements
there are five Medusa fibres assigned for simultaneous arc
calibrations.  These are distributed evenly across the CCD and (in the
absence of information to the contrary at the time) were left enabled
for the Galactic observations.  Inspection of these arc spectra showed
no significant wavelength shifts (for a given central wavelength) in
comparison to the daytime calibrations.  However, these checks did
highlight the overspill of the simultaneous arc spectra into those
from adjacent fibres on the CCD.  Several of our science spectra
contain arc-line contamination from this overspill, although the
contamination is generally weak and in the majority of cases limited
to the red (HR14, \lam6515) wavelength setting.

\subsection{Comparison with pipeline reductions}
During the course of this survey, the Giraffe Base-Line Data Reduction
Software (girBLDRS) underwent significant development at Observatoire
de Gen$\grave{\rm e}$ve \citep[see][ for full details]{girbldrs}; this
pipeline has been used for reduction of the Magellanic Cloud
observations (Evans et al, in preparation).

Prior to reduction of the Magellanic Cloud data, we re-reduced the
observations in NGC\,3293 using girBLDRS (v.1.10) for comparison with
the IRAF reductions.  In brief, we used the calibration data observed
at the time of our science frames to update those released with
girBLDRS, and then used the `extract' pipe.  The pipeline default is
an optimal extraction of each fibre, however this option relies on
correction of the localisation of the fibres on the CCD.  The master
localisation is obtained using the flat-field frames, but the
correction employs the simultaneous arc spectra.  In anticipation of
the fact that the simultaneous arcs were disabled for the majority of
our FLAMES observations in the Magellanic Clouds (because of the
overspill discussed in Section~\ref{arc_probs}), the simpler summed
extraction method was used for our comparisons.  After extraction with
girBLDRS, a combined sky spectrum was subtracted from each target using
{\sc DIPSO}, employing the same methods as those used in the IRAF
reductions.  These re-reduced data were rectified using the same
scripts as before and the final spectra are in excellent agreement.

\subsection{Comparison with longslit spectroscopy}

We have conventional longslit spectroscopy of the majority of our
O-type targets in NGC\,6611 (see Table \ref{6611}).  These were
observed as part of separate programmes at the 4.2-m William Hershel
Telescope (WHT) on 2003 June 15-17 and October 16-17, using the
Intermediate-dispersion Spectroscopic and Imaging System (ISIS) with
1200B and 1200R gratings.  Observations were made at three central
wavelength settings in both arms of ISIS, giving continuous spectral
coverage of 3800-5100\,\AA, with additional coverage of
6200-6800\,\AA, at $R \simeq$ 7000.  The basic properties of the ISIS 
observations are summarized in Table~\ref{instrumentation}, in which 
they are compared with those from FLAMES-Giraffe and FEROS.

\begin{table}
\caption[]{Overview of the wavelength coverage and resolution obtained
from the different instrumentation employed in the current study.}
\label{instrumentation}
\begin{center}
\begin{tabular}{lcc}
\hline\hline
Spectrograph   & Wavelength coverage (\AA) &     $R$       \\
\hline
FLAMES-Giraffe & 3850-4755, 6380-6620      & 20,850-29,600 \\
FEROS          & 3600-9200                 & 48,000        \\
WHT-ISIS       & 3800-5100, 6200-6800      & 7,000         \\
\hline
\end{tabular}
\end{center}
\end{table}

In general we find excellent agreement between the FLAMES-Giraffe and
ISIS spectra (aside from detection of binarity, see Section
\ref{binaries}).  Indeed, minor differences in the final spectra arise
in the far-blue region, not because of problems with the FLAMES
spectra but from residual features from the ISIS dichroic used to
simultaneously observe in both blue and red wavelength regions.  Of
particular note is that the equivalent widths of important diagnostic
lines (such as He~\1 \lam4471, He~\2 \lam4686 and H$\gamma$) agree,
adding confidence to the multi-fibre FLAMES data.

\begin{figure*}
\begin{center}
\includegraphics{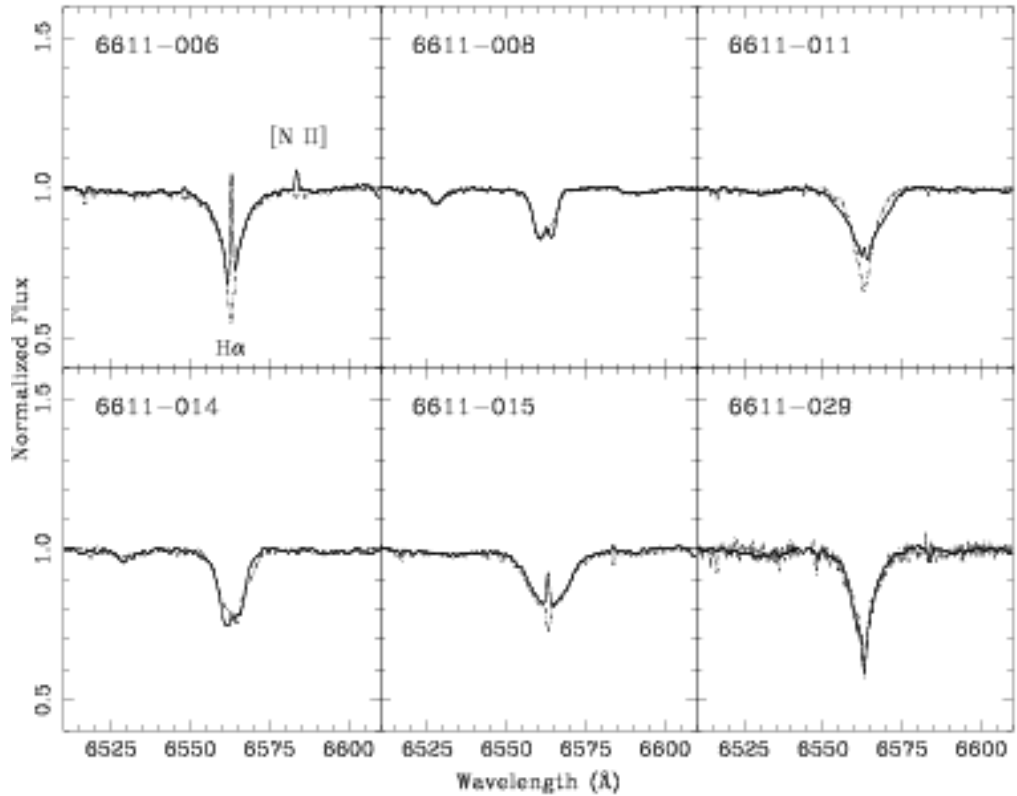}
\caption{Comparison of sky-subtracted FLAMES-Giraffe (solid line) 
and WHT-ISIS (dot-dashed line) H$\alpha$ profiles, in which the FLAMES
data have been smoothed to the effective resolution of the ISIS
spectra.  The nebular subtraction in the FLAMES spectra is not perfect
but its contribution is resolved.  Note the lower signal-to-noise ratio in
the ISIS spectrum of 6611-029, and the differences in 6611-011 and
6611-014, explained by binarity (see Section~\ref{binaries}).
\label{halpha}}
\end{center}
\end{figure*}

In Figure \ref{halpha} we compare the H$\alpha$ Balmer line profiles
for the six stars in NGC\,6611 observed with both FLAMES-Giraffe and
WHT-ISIS; the FLAMES spectra have been degraded to the effective
resolution of the WHT-ISIS data to ensure a more meaningful
comparison.  As one would expect from multi-fibre observations, the
nebular subtraction is not perfect in the FLAMES spectra.  However, the
nebular component is easily resolved and the wings of the profiles are
in good agreement with the longslit spectroscopy, and will still offer
strong constraints on the physical parameters.  The differences in
Figure~\ref{halpha} for 6611-011 and 6611-014 are accounted for by
binarity, further discussed in Section~\ref{binaries}.

\section{Spectral Classification}
The FLAMES-Giraffe spectra of our Galactic targets were classified in the MK
system \citep{mkk} by visual inspection.  The Giraffe spectra have a
much greater resolution than those traditionally used in spectral
classification, so the data were degraded to an effective
resolution of 1.5\,\AA~and then classified according to published
morphological precepts.  Work is underway by other groups
\citep[e.g. ][]{bcj03} to compile high-resolution observations of
spectral standards and our survey will significantly add to such
efforts.  Furthermore, some of our stars have robust spectral types in the
literature and so also serve as internal standards for the current dataset.  The
classification criteria applied to each spectrum are summarized below,
points of interest and spectral peculiarities are also discussed.
Following initial classification by CJE, the data were reinspected to
ensure internal consistency and, if necessary, minor revisions of
spectral type were made.  A subset of these data were independently
classified by DJL to provide an external check on the spectral types;
agreement to within one spectral subtype was found in all cases.  The
distribution of observed stars by spectral type is summarised in Table
\ref{sp_sum}.

\begin{table}
\caption{Total numbers of stars in each cluster by spectral type (incorporating
both the FLAMES and FEROS observations).}
\label{sp_sum}
\begin{center}
\begin{tabular}{lccccccc}
\hline\hline
Field     & \multicolumn{6}{c}{Spectral Type}        & Total \\
          &  O  & B$<$5 & B$\ge$5 &  A  &  F   & GK  &       \\
\hline
NGC\,3293 & $-$ &  48   &  51     & 24  &  3   & $-$ &  126  \\ 
NGC\,4755 & $-$ &  54   &  44     &  9  &  1   & $-$ &  108  \\
NGC\,6611 &  13 &  28   &  12     & 10  & 10   & 12  &   85  \\
\hline
\end{tabular}
\end{center}
\end{table}

\subsection{O-type spectra}
\begin{figure*}
\begin{center}
\includegraphics{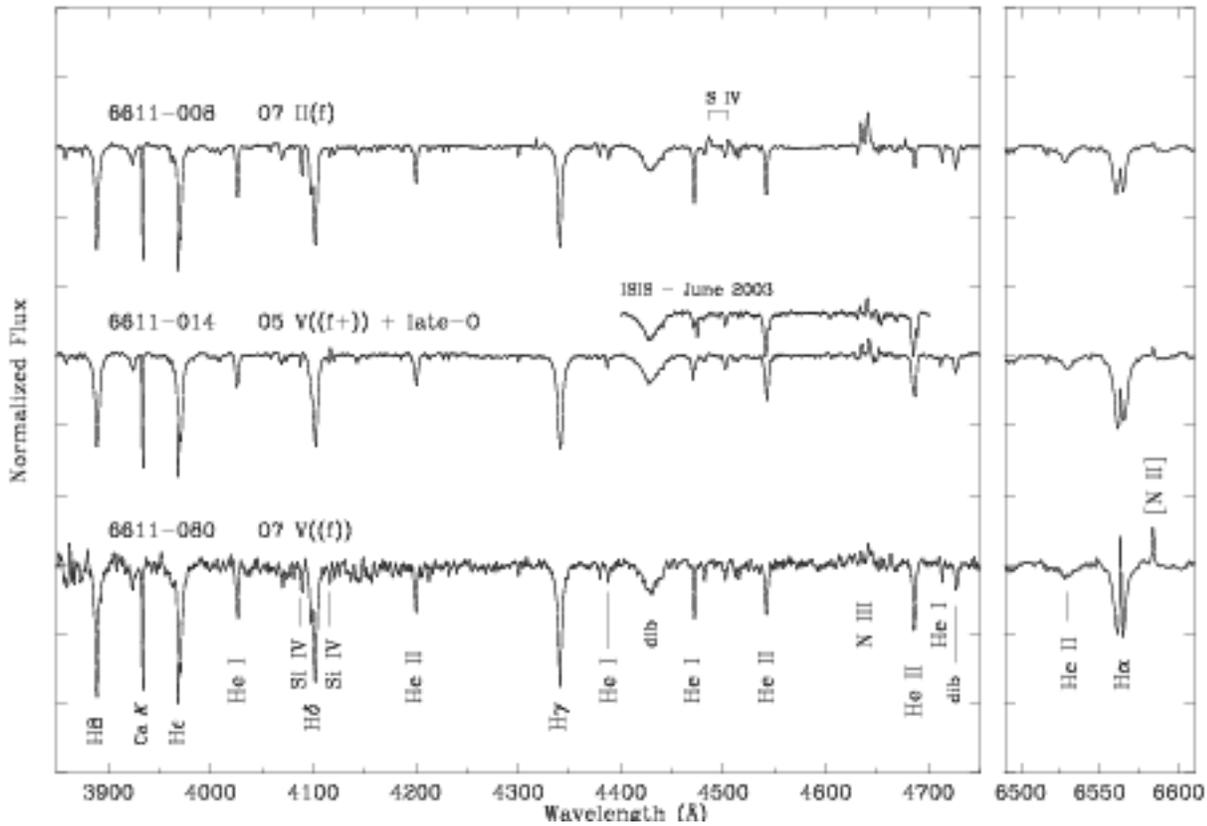}
\caption[]{Illustrative FLAMES O-type spectra in NGC\,6611.  For display purposes the 
spectra have been smoothed and rebinned to a resolution of
1.5\,\AA~FWHM and are offset by 0.75 continuum units.  A section of
the longslit ISIS spectrum of 6611-014 is also shown, illustrating its
binarity (see Section \ref{binaries}).  The sky spectrum has been
subtracted from the blue-region data but, to highlight the nebular
contribution, it has not been removed in the red.  The spectral lines
identified in 6611-080 are, from left to right by species,
H~{\footnotesize I} \lam\lam3889 (H8), 3970 (H$\epsilon$), 4102
(H$\delta$), 4340 (H$\gamma$), 6563 (H$\alpha$); He~{\footnotesize I}
\lam\lam4026, 4388, 4471, 4713; He~{\footnotesize II} \lam\lam4200,
4541, 4686, 6527; Si~{\footnotesize IV}
\lam\lam4089, 4116; N~{\footnotesize III} \lam\lam4634-40-42; and the
interstellar Ca~$K$ \lam3933 line.  Also marked are the diffuse
interstellar bands at \lam4430, 4726 and nebular
N~{\footnotesize II} forbidden emission at \lam6583.  The
S~{\footnotesize IV} \lam\lam4486, 4504 emission lines are marked in
6611-008.}
\label{ostars}
\end{center}
\end{figure*}

The O-type spectra were classified using the digital atlas of
\citet{wf90} as the principal reference; the primary classification
criterion is the ratio of He~\1 to He~\2 lines.  A total of 13 O-type
stars were observed, all of which are in NGC\,6611.  The standard
suffixes are used to indicate the intensity of N~\3 emission and the
behaviour of He~\2 \lam4686 (see \citeauthor{wf90} for further
details).  FLAMES-Giraffe spectra of three O-type stars in NGC\,6611
are shown in Figure \ref{ostars}; note that 6611-080 is one of the
faintest targets in the Galactic sample and yet the data are of
sufficient quality to permit detailed analysis in a future study.

\subsection{B-type spectra}
\label{btypes}
The observed B-type spectra were classified from comparisons with the
standards in the \citet{wf90} atlas, which extends as late as B3 for
dwarfs.  These standards were complemented by unpublished
observations of B-type dwarfs using the Intermediate Dispersion
Spectrograph (IDS), at the 2.5-m Isaac Newton Telescope (INT) in 1999
December, and with FEROS in 2004 July.  These additional standards are
summarized in Table~\ref{bstds}.

Some of the Giraffe spectra are assigned the non-unique spectral type
of B6-7 (see Tables \ref{3293} and \ref{4755}).  Examples of B6 and
B7-type supergiants were presented by \citet{ldf92}, with subtle
variations seen in the metallic spectra as one progresses from B5
through to B8.  The FLAMES spectra however are largely of
less-luminous objects and such subtleties are not seen because of the
intrinsically weaker metallic lines in non-supergiants.  Additionally,
rotational velocities are typically larger in non-supergiant B-type
stars, which further confuses the issue.  As a result, here we
classify those spectra that are clearly between the observed B5 and B8
standards as `B6-7'.

\begin{table}
\caption[]{Summary of B-type classification standards in addition to those
from \citet{wf90}.}
\label{bstds}
\begin{center}
\begin{tabular}{llll}
\hline\hline
Star      & Sp. Type & Telecope & Reference \\ 
\hline
HD 198781 & B0.5 V   & INT      & \citet[MCW 950; ][]{mcw55}\\
HD 24131  & B1 V     & INT      & \citet[MCW 251; ][]{mwc53}\\
HD 215191 & B1.5 V   & INT      & \citet{l68}  \\ 
HD 212978 & B2 V     & INT      & \citet{bm53} \\
HD 182180 & B2 Vn    & ESO-2.2  & \citet{hgs69} \\ 
HD 23625  & B2.5 V   & INT      & \citet{l68}  \\
HD 20365  & B3 V     & INT      & \citet{rm50} \\
HD 25558  & B3 V     & INT      & \citet{b56}  \\
HD 19268  & B5 V     & INT      & \citet{l68}  \\
HD 34233  & B5 V     & INT      & \citet{l68}  \\
HD 58715  & B8 V     & INT      & \citet{mkk}  \\
HD 224112 & B8 V     & ESO-2.2  & \citet{hgs69}\\ 
HD 107696 & B9 V     & ESO-2.2  & \citet{hgs69} \\ 
\hline
\end{tabular}
\end{center}
\end{table}

The width of the lines in the B-type spectra offer
evidence of a wide range of rotational velocities, some are
particularly sharp-lined, whilst others are significantly broadened.  For now
we limit ourselves to simply adding the conventional `n' suffix to
those B-type spectra which display the most significant broadening, employed
here when the He~\1 \lam4471 and Mg~\2 \lam4481 lines are strongly blended.
A thorough analysis of the distribution of rotational velocities will be
presented elsewhere (Lee et al., in preparation).

\subsubsection{Luminosity classification}
Luminosity types were assigned to the B-type spectra using
measurements of the H$\gamma$ equivalent widths and the calibrations
of \citet{bc74}.  In some cases the measured widths are mid-way
between the published values for class III and V objects, resulting in
an adopted class of `III-V'.

\citet{bc74} largely confirmed the earlier results of \citet{p65} for 
B-type spectra. Their (photographic) work remains the most useful and
reliable source of such calibrations, although some less wide-ranging
studies (primarly concerned with absolute magnitudes of supergiants,
of use in consideration of extragalactic distances) have made use of
more modern detectors \citep[e.g.][]{hwy86}.  We envisage that with
completion of our survey of Galactic and Magellanic Cloud targets we
will be in a position to revisit these relations with an unprecedented
(and fully homogenous) dataset.  Indeed, two of the clusters observed
by \citet{bc74} were NGC\,3293 and 4755; pending a new study we do not
explore these issues further in the current work.

\subsection{A-type spectra}
Digital observations of A-type spectra were presented by \citet{eh03}.
Their principal classification criterion uses the intensity of the
Ca~$K$ line compared to the H$\epsilon$/Ca~$H$ blend.  Although their
work was principally motivated by studies of A-type supergiants, the
adopted scheme was found to be largely insensitive to luminosity
effects (e.g. their Figure~4).  However, reliance on the calcium line
ratio is not without problems (see discussion by \citeauthor{eh03}),
and with the high quality of the FLAMES data we are also able to give
consideration to the metallic features in the spectra, which increase
as one progresses to later-types.  Thus, the observed A-type stars
were classified using the criteria of \citet{eh03}, with secondary
consideration to the intensity of the metal-lines.  Luminosity types
were allocated using the H$\gamma$ equivalent-width criteria from
\citet{eh04}

\subsection{Later-type spectra}
As a result of the significant and variable reddening in the direction
of NGC\,6611, our target selection methods yielded numerous (22 of 73)
targets with spectral types of F0 or later.  Without prior
spectroscopy to make informed decisions in target selection, this is
not a surprising result.  It is not our intention to analyse these
stars in the future, but they are included here for completeness.  The
classification bins applied to these later-type spectra are largely
those described by \citet{eh03}, with additional reference to the
useful on-line digital spectral atlas of
R. O. Gray\footnote{http://nedwww.ipac.caltech.edu/level5/Gray/Gray\underline{
}contents.html}, for those in the region G5-K2.

\subsection{H$\alpha$ morphology}
Useful as a diagnostic of stellar winds and Be-type stars, when of
particular note the morphology of the H$\alpha$ profiles of our targets is
summarized in the final column of Tables \ref{3293}, \ref{4755} and
\ref{6611}.  The majority of the sample display H$\alpha$ in absorption,
with many of these also showing superimposed (narrow) nebular
emission.  Such stars would be best described as `abs$+$em' but are
not explicitly noted in the tables -- exceptions to this are
3293-001 and 4755-003 in which the emission is broader and more likely
stellar in origin.  The notation employed to describe the other
H$\alpha$ profiles is: `broad em' for broad emission; `twin' for
twin-peaked emission (usually synonomous with disk-like features);
`abs$+$twin' for those spectra in which twin-peaked emission is seen
superimposed on an absorption profile; and `P Cyg' for stars showing a
P Cygni emission profile.  

\subsubsection{Emission line spectra}
A small number of emission line objects were observed in our Galactic
fields, the majority of which can be categorized as normal Be-type
spectra.  Three spectra (namely 3293-022, 3293-045, and 4755-014) have
strong twin-peaked H$\alpha$ emission profiles; five (3293-027,
3293-040, 4755-038, 4755-050 and 4755-057) have twin-peaked emission
visible in the wings of the H$\alpha$ lines, superimposed on
absorption profiles; and four (3293-011, 4755-018, 6611-010 and
6611-028) display broad, single-peaked emission features.  These have
all been classified as Be or Ae stars.  Broad H$\alpha$ emission is
also seen in 6611-022, which is discussed further in Section
\ref{oddspectra}.

Further caution regarding the problems of nebular subtraction is 
prompted by the significantly large number of Be/Ae objects in 
NGC\,6611 reported by \citet{hmsm}.  From observations with a
slitless, grism spectrograph, \citet{hd01} attribute most of the 
emission features as due to nebular contamination,
rather than from intrinsic stellar phenomena.  For instance, the
H$\alpha$ emission in four targets reported as Be-type by
\citeauthor{hmsm} is best described here as `absorption with narrow
(presumably nebular) emission' (6611-020, 064, 066 and 068).  Each of
these is reported by \citeauthor{hd01} as having H$\alpha$ absorption
profiles (albeit their data are at relatively low resolution); these
results reinforce our cautious approach to employing the Be-notation

\subsection{Peculiar spectra}
\label{oddspectra}
\begin{figure*}
\begin{center}
\includegraphics{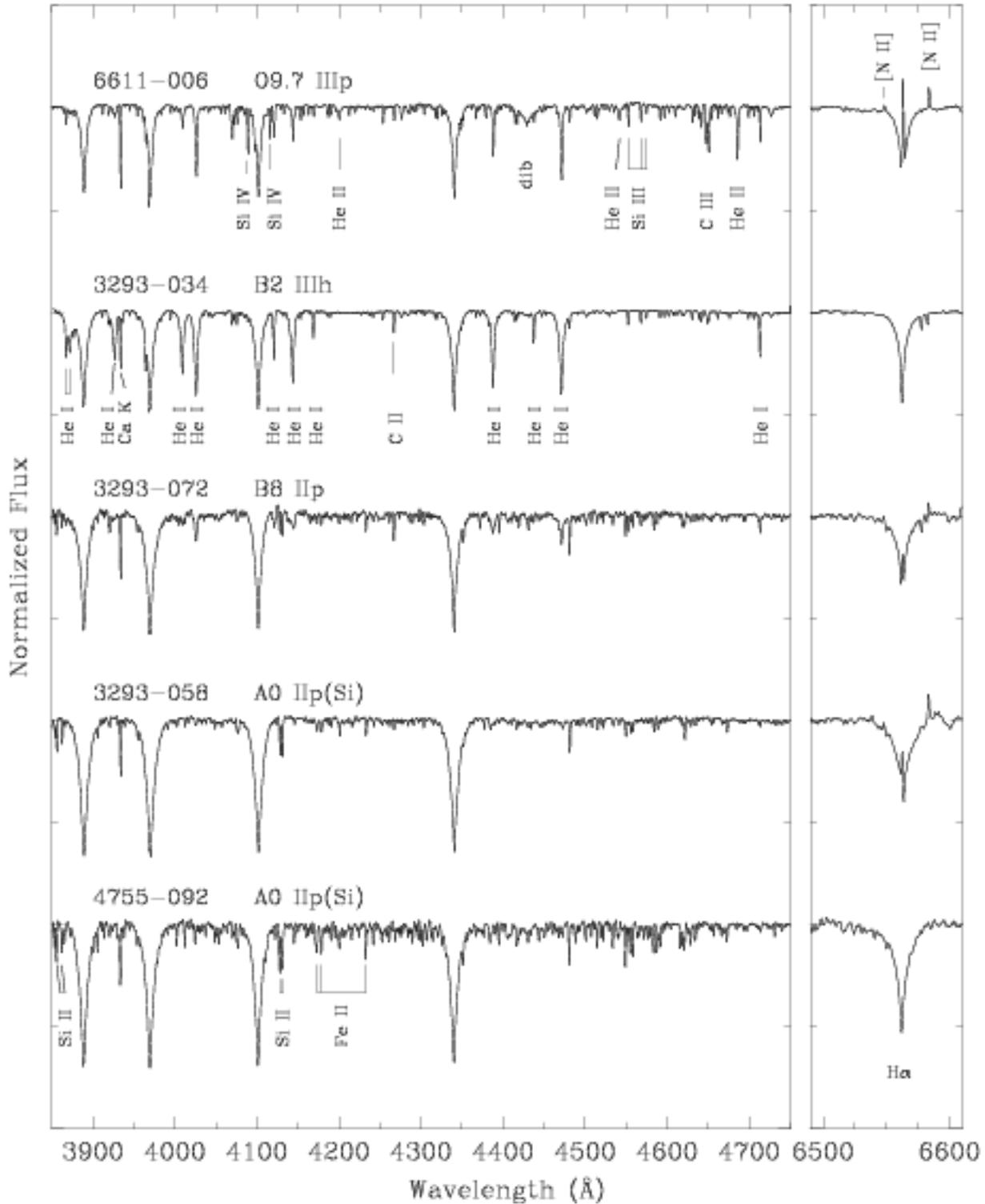}
\caption[]{FLAMES-Giraffe spectra of particular interest.
The spectral lines identified in 6611-006 are, from left to right
by species, He~{\footnotesize II} \lam\lam4200, 4541, 4686; 
Si~{\footnotesize III} \lam\lam4553-68-75; Si~{\footnotesize IV} 
\lam\lam4089, 4116; the C~{\footnotesize III} $+$ O~{\footnotesize II} 
blend at \lam4650; the N~{\footnotesize II} forbidden emission lines
at \lam\lam6548, 6583; and the diffuse interstellar band at \lam4430.
The lines identified in 3293-034 are, He~{\footnotesize I} 
\lam\lam3867$+$3872, 3926, 4009, 4026, 4121, 4144, 4169, 4388, 4438, 
4471, 4713; C~{\footnotesize II} \lam4267; and the interstellar
Ca~$K$ line at \lam3933.  
Lastly, the metallic lines identified in 4755-092 are Si~{\footnotesize II}
\lam\lam3856, 3862, 4128-30; and Fe~{\footnotesize II} 
\lam\lam4173-78, 4233.  For display purposes the spectra have been smoothed
and rebinned to a resolution of 1.5\,\AA~FWHM and are offset by 1 continuum unit.}
\label{oddstars}
\end{center}
\end{figure*}

A number of peculiarities were noted in the process of classification, 
which are now discussed in turn:
\begin{description}
\item {\bf 3293-034:}  The hydrogen lines and metallic spectrum are consistent 
with a spectral type of B2, but the He~\1 lines are significantly enhanced 
(see Figure \ref{oddstars}) and it is therefore classified as B2 IIIh
\citep[cf. ][]{w83}.

\item {\bf 3293-058:}  Significantly enhanced Si~\2 \lam4128-32 is seen in this
spectrum, which is otherwise that of a normal A0-type (see Figure
\ref{oddstars}).  Such behaviour in the silicon lines is a well
documented feature of Ap-type spectra \citep[e.g. ][]{m33} and the
spectrum is classified as A0 IIp (Si).

\item {\bf 3293-072:}  This star has a rich metallic spectrum and from 
the relative ratios of He~\1/Mg~\2 and the intensity of the Si~\2
lines is classified here as B8~II.  However, some of the He~\1 lines
appear peculiar in that they appear more broadened and asymmetric
than the relatively sharp metal lines (see Figure \ref{oddstars}).
Inspection of the raw data (i.e., unsmoothed) in velocity space
revealed no consistent component for the He~\1 lines; although the
plausible explanation for their peculiar appearance remains a hotter,
though less luminous companion.  In the absence of further information
at the current time, we classify the spectrum as B8~IIp.

\item {\bf 4755-092 \& 4755-104:}
These two spectra also display enhanced Si~\2 and are classified as A0
IIp (Si).  In comparison to 3293-058, other metallic species also
appear slightly more enhanced in these stars, e.g. lines from Fe~\2.
The spectrum of 4755-092 is shown in Figure \ref{oddstars}. 

\item {\bf 6611-006:}
This star has been classified by a number of authors as O9.5 V
\citep{hm69,hmsm,bmn99} but from the FLAMES spectrum
it is clear that the strength of the Si~\3 triplet at
\lam4552-68-75 is inconsistent with such a type.  
This star was subsequently observed with FEROS, with no obvious
evidence of binarity found.  Taking into account the effects of
metallicity, the spectrum is similar to that of AzV~170 in the SMC
\citep{wal00} and we classify it here as O9.7 IIIp.  There is no
formal definition of the O9.7 type below class II \citep{w72} and a
marginally later type of B0~IV \citep[cf. the standards from][]{wf90}
might be more fitting; there is also a rich spectrum of carbon and
oxygen lines.  Pending further examination of the spectral standards
in this domain, for now we employ the `p' suffix to denote
peculiarity.  To place this particular object in its broader context,
it is the principal star seen in the now widely circulated {\it Hubble
Space Telescope (HST)} Wide Field and Planetary Camera 2 (WFPC2) image
of the `pillars' in the Eagle Nebula (Hester \& Scowen, 1995, Arizona
State University/NASA).

\item {\bf 6611-022:}
The spectrum of 6611-022 (Walker \#235) is rich in Fe~\2 emission features and
displays significant, broad H$\alpha$ emission.  It is classified here
\citep[as by other authors e.g., ][]{hd01} as Herbig Be, following the
identification of such stars by \citet{h60}.  At the high resolution
of our data we resolve two distinct peaks in each emission feature, strongly
indicative of some form of rotating disk or shell around the star.

\end{description}

\subsection{Interstellar absorption lines}
\label{3888}
Strong interstellar Ca~\2 $H$ and $K$ lines at \lam3933 and
\lam3968 are observed towards all three clusters, although blended
with stellar features for the cooler targets. For the two older clusters
(NGC\,3293 and NGC\,4755) the absorption is not resolved into separate
components and probably arises mainly in the warm diffuse interstellar
medium; no other interstellar lines are apparent although
it should be noted that the spectral region containing the Na~\1
doublet was not observed.

The interstellar spectra for the NGC\,6611 targets are far richer,
likely reflecting the substantial circumstellar and interstellar
material associated with this young cluster.  For example,
\citet{the90}, \citet{hmsm} and \citet{ys01} have all discussed both the variable
extinction across the cluster and the non-standard reddening law.
Furthermore, \citet{welsh84} found evidence for high velocity
interstellar components in IUE spectra.  He interpreted these as
arising from a shock-front (that is driven by stellar winds)
interacting with the ambient neutral interstellar gas.  In the
NGC\,6611 FLAMES-Giraffe spectra, the Ca~\2 lines show significant
spatial variations - with a single strong component (which could be
unresolved multiple components), complemented by up to four additional
high velocity components.  The strong component probably again
reflects the warm diffuse interstellar medium, whilst the other
features are likely to be spatially associated with the cluster. 

Many sightlines show molecular absorption lines (with significant
variations in the line strengths) from CH (\lam4300), CH$^{+}$
(\lam4232), and CN (\lam3875), implying the presence of a cold
molecular cloud.  In some targets the Ca~\1 \lam4227 line is also
observed, further evidence of a dense, optically-thick cloud.
Finally, He~\1 absorption at \lam3889 (from the 2s$^3S\rightarrow$
3p$^3P$\, transition) is observed towards some objects, but with large
spatial variations.  This was first observed by
\citet{w37} and arises in an ionized plasma in which the helium ions
are recombining to a metastable triplet state \citep[e.g.][]{o97}.  
The source of the ionization in this instance is presumably the strong X-ray
continuum emitted by the O-type stars in the cluster.

\subsection{Comparisons with previous spectroscopy}
\label{previous}
In general there is good agreement between our classifications and
those extant in the literature, with many of the fainter stars now
receiving more refined classifications, as one would expect from the
new high-quality data.  In Table 12\footnote{Available in the online
version of the journal} we give previous
spectral types for our target stars.  Published classifications are
included from: \citet{b54}; \citet{bmn99}; \citet{cl74}; \citet{f3293}; 
\citet{f4755}; \citet{her60}; \citet{hmsm}; \citet{hgs69}; \citet{hm69}; 
\citet{h56}; \citet{mcw55}; \citet{s70}; \citet{turn80}; \citet{w61}; \citet{w72}; 
\citet{w73}; \citet{w76}; and \citet{w82}.  We omit the spectral 
types in NGC~6611 from \citet{dW97} as they are largely based on
photometric methods, and are thus `low weight' classifications.

Many of the targets in NGC\,3293 (92 of 126 stars) and NGC\,4755 (86
of 108 stars) do not have previous classifications from the sources
compiled in Table \ref{previous}.  Even in NGC 6611, in which the
spectral content has been well explored in the past \citep[most
significantly by][]{hmsm}, we provide classifications for 31 stars
without previous types in the literature.  Our main science drivers do
not require spectroscopic completeness but, to be thorough, we note
that due to observational constraints (primarily the fact that only
one field configuration was observed) a small number of B-type stars
with published spectral types from \citet{hmsm} are not included in
the current survey.

\section{Stellar radial velocities}
\label{rvs}

Radial velocities ($v{_{\rm r}}$) of our targets were measured from
the Doppler shift of the H\,$\gamma$, H$\delta$, H$\epsilon$, and H8
(H$\zeta$) Balmer lines.  This relatively simple approach (compared to
using multiple metallic lines) was adopted to ensure internal
consistency over the wide range of observed spectral types.  The line
centres were measured by Gaussian fits to the inner region of the
profile, or by manual estimates of the position of minimum intensity.
Only the inner part of the Balmer lines ($\pm$3 \AA\/ of the centre)
were used for the Gaussian fits because of the effect of linear Stark
broadening on the wings; the whole profile is poorly fit by a
Gaussian function but, in general, the inner part is successfully matched.
This approach worked well for symmetric profiles and was also
particularly effective when weak nebular emission features were
present in the line centres, because the fits are relatively
unaffected.  In cases where the lines were very sharp or asymmetric
this approach was not successful and the line centres were determined
from manual measurements.  In the double-lined binary 4755-024, two
Gaussian profiles were fitted, with the quoted value
being that of the primary.

\begin{figure*}
\begin{center}
\caption{Distribution of stellar radial velocities.  The left-column plots show the 
radial velocity of each star, with the 1-sigma scatter shown in the
lower panel for each cluster.  The right-column plots are the binned
histograms for each cluster, with a 10 \kms\/ bin size.}
\label{rvfig}
\includegraphics{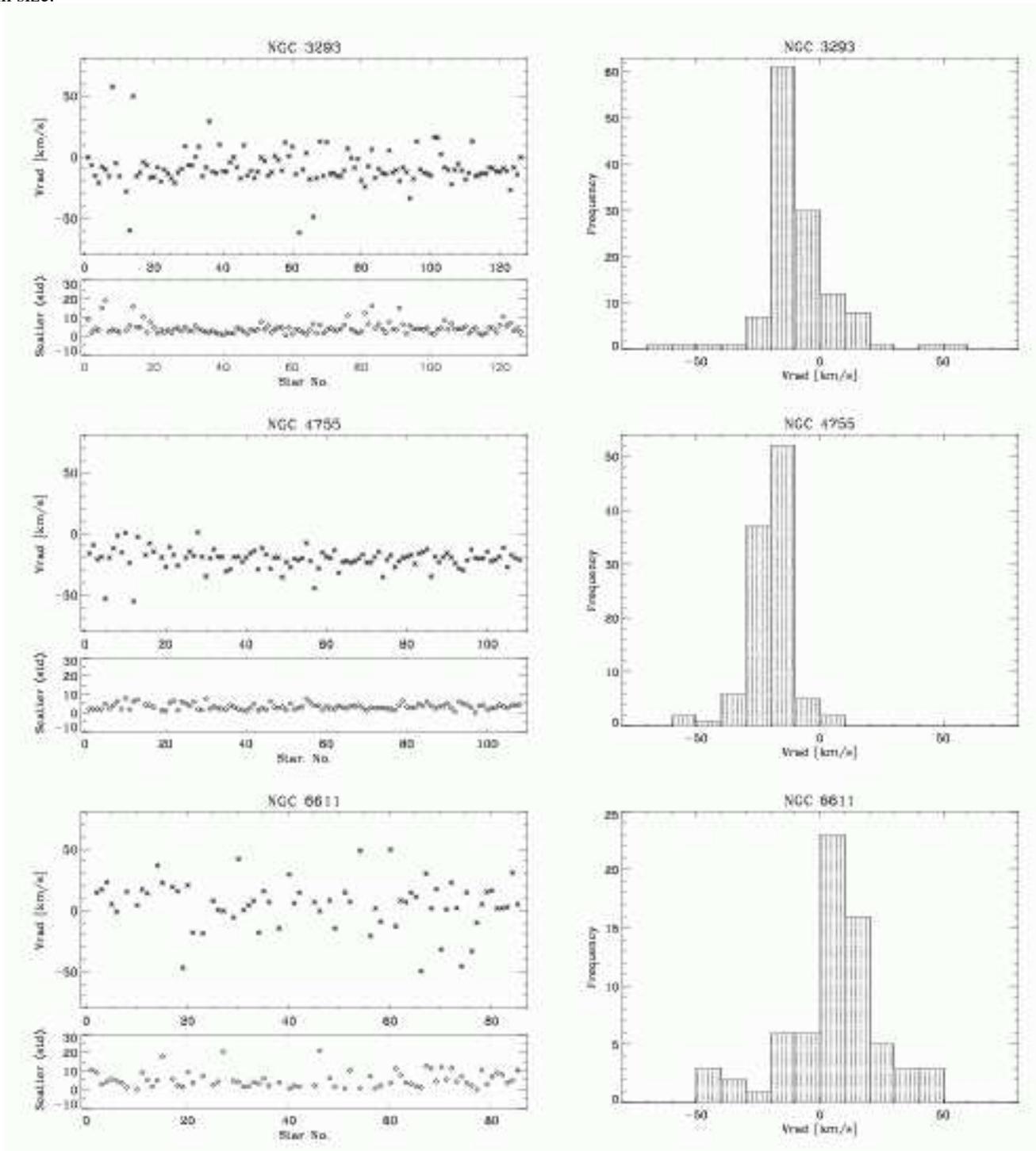}
\end{center}
\end{figure*}

Heliocentric radial velocities are included in Tables \ref{3293},
\ref{4755}, and \ref{6611}.  The values are plotted for the stars
in each cluster in the left-column of Figure~\ref{rvfig};
the upper panels show the mean velocity for each star, with the
standard deviation ($\delta v_{\rm r}$) of the individual measurements
shown in the lower panels.  The radial velocity distribution for each
cluster is shown in the right-column plots of Figure \ref{rvfig}.

There are a number of specific comments regarding the radial velocity
measurements.  In the O- and B-type spectra the interstellar Ca\,$H$
\lam3967 line is easily resolved and was ignored when fitting the
H$\epsilon$ line.  In contrast, in the A-type spectra the stellar
component of the Ca\,$H$ line becomes increasingly strong, and the
H\,$\epsilon$ line was excluded from calculation of the mean velocity.
Similarly, as discussed in Section \ref{3888}, some stars in
NGC\,6611 display sharp and strong He~\1 absorption close to the
centre of the H8 line; this line was not included in calculation of the
mean velocities for any of the NGC\,6611 stars.  For the G- and K-type
spectra in NGC\,6611 the quoted velocities are the means of manual
measurements of several Fe~\1 lines (\lam\lam4071.7, 4528.6, 4602.9)
and the Cr~\1 \lam4254.3 line.  

The mean velocity and its associated dispersion for each cluster is given in Table 
\ref{rvresults}.  In calculation of the mean velocities, their dispersion
around the mean, and the mean standard deviation of the measurements
within each stellar spectrum, it is appropriate to exclude cluster
non-members.  In NGC\,3293 and NGC\,4755 we consider stars with
spectral types of B8 or later as probable non-members.  Given the
imposed faint magnitude limit applied in target selection this is
reasonable -- indeed any B8 or later-type stars observed are often
classified as giants and are therefore likely to be non-members of
these relatively young clusters (or at least non-coeval with the
observed upper main-seqeunce).  Following similar arguments in
NGC\,6611, we do not include stars of type B5 or later.

\begin{table}
\caption[]{The mean radial velocity of the clusters $<v_{\rm r}>$, with their dispersions
$\sigma_{v}$, and the mean error in the individual measurements
$<\delta v_{\rm r}>$.  The number of stars, $n$, used for each cluster
(after removal of outliers) is also given.}
\label{rvresults}
\begin{center}
\begin{tabular}{lcccc}
\hline\hline
Cluster          &  $<v_{\rm r}> \pm \sigma_{v}$ & $<\delta v_{\rm r}>$  &  $n$ \\
                 &         [\kms]            &    [\kms]         &      \\
\hline
NGC6611          & $+$10 $\pm$ 8             &   5               &  33  \\
NGC3293          & $-$12 $\pm$ 5             &   4               &  70  \\ 
NGC4755          & $-$20 $\pm$ 5             &   3               &  56  \\
NGC4755 (bright) & $-$14 $\pm$ 8             &   3               &  16  \\
NGC4755 (faint)  & $-$21 $\pm$ 5             &   3               &  45  \\
\hline
\end{tabular}
\end{center}
\end{table}

The mean of the standard deviations of the individual radial
velocities was calculated to give an estimate of the expected
uncertainty (referred to hereafter as $<\delta v_{\rm r}>$), and this can
be directly compared to the velocity dispersion measurements in the
clusters to determine if we are measuring statistically significant
velocity dispersions.  These values (i.e. $<\delta v_{\rm r}>$), the mean
radial velocity of each cluster ($<v_{\rm r}>$) and the dispersion
($\sigma_{v}$) are also affected by the stars with variable radial
velocities, some of which are the outliers in the plots in Figure
\ref{rvfig}.  To exclude these we employed a consistent method.  The
mean ($<v_{\rm r}>$) and standard deviation ($\sigma_{v}$) of the velocity
distributions were calculated, and any star lying more than
2$\sigma_{v}$ from $<v_{\rm r}>$ was rejected and the process
repeated. This iteration converged quickly and resulted in clear
outliers being removed. The same method was used to determine the
intrinsic mean error in our analysis $<\delta v_{\rm r}>$ in each
cluster. The values are given in Table \ref{rvresults}, together with
the final number of stars used, $n$.  For each of the clusters the
measured velocity dispersion is less than 2$<\delta v_{\rm r}>$, i.e., we
do not see any strong evidence for detecting real velocity
dispersions. The measurement errors could possibly be reduced using
more sophisticated velocity measurement techniques, such as
cross-correlation with model spectra, however the rationale for that
is beyond the scope of this paper.  Note that the mean standard
deviation of the individual measurements $<\delta v_{\rm r}>$ ranges from
3-5 \kms, approximately the one-pixel resolution of the spectra.
Lastly, significant nebular emission or broad, asymmetric profiles
made precise location of the line centre difficult in a number of
stars; these are flagged with the usual `:' identifier in the tables
to indicate a greater uncertainty ($\sim$20 \kms) than for the other
values.

Figure \ref{rvfig} shows that there is a tendency for the
bright cluster members in NGC\,4755 to have a smaller approach velocity $v_{\rm r}$
than the fainter ones. This was suggested by \citet{f4755}, although
he saw this effect when he combined data from three clusters. We see
no evidence in NGC\,3293 and NGC\,6611 for the effect, however in
NGC\,4755 the $<v_{\rm r}>$ values for the bright sample (star numbers
$\leq$20, or $V<$10) is indeed significantly different from the faint
sample (star numbers $>$20, or $V>$10), see Table \ref{rvresults}.
Whether this is physically meaningful remains unclear.

\section{Binarity}
\label{binaries}
The repeated observations of the NGC\,6611 field offer an opportunity
to check for single-lined spectroscopic binaries.  We find evidence of
two stars with companions from this method, 6611-030 and 6611-068.
Similarly, the July 18 spectrum of 6611-007 appears to show a
relatively sharp-lined early B-type star, however, the July 15
observation reveals it as a double-lined spectroscopic binary.  A B0.5
V spectral type is assigned to one component because of weak He~\2
\lam4686 absorption; the other component is also early B-type,
approximately B1.

The two FLAMES observations of 6611-014 display some small profile
asymmetries but otherwise appear largely unremarkable.  However,
inspection of the ISIS spectrum of this star reveals it to be a
double-lined binary.  From the FLAMES spectra we had initially
classified this object as O6 V((f$+$)), the suffix indicating the
N~\3 emission and evidence of Si~\4 emission at \lam4116.  From
the ISIS spectrum the He~\1 \lam4471 line of the hotter component
appears weaker than for an O6 and we assign a spectral type of O5
V((f$+$)).  The cooler component appears to be that of a late O-type
with He~\1
\lam4471 stronger than the components of the He~\2 lines at \lam4200
and 4541.  This confirms the suggestion of \citet{bmn99} that 6611-014
(W175) comprises two O-type stars \citep[note that a third, relatively
low-mass component was reported by][]{ds01}.

Two further stars in NGC\,6611 were identified as binaries by 
\citet{bmn99}, namely 6611-001 (W412) and 6611-003 (W197).  
They found that the former is a short period ($\sim$4 days)
single-lined binary, but the FLAMES-Giraffe and ISIS spectra are in
good agreement and thus offer no additional insight.  In the case of
6611-003 (initially classified here as O7) \citeauthor{bmn99} noted
the presence of weak Si~\3 absorption that is inconsistent with a mid
O-type classification, suggesting a cooler companion.  Indeed, shifts
of up to 100 \kms are seen in the Si~\3 lines between the ISIS and
FEROS spectra, similarly with the C~\3 blend at \lam4650.  The
intensity of the \lam4200 and \lam4542 He~\2 lines is invariant, but
the \lam4686 line is weaker in the ISIS spectrum, reinforcing the
suggestion by \citeauthor{bmn99} of an approximately B0-type secondary
-- such a star would contribute to \lam4686, but have less of an
impact on the other He~\2 lines.  The presence of such a secondary
could justify revision of the classification of the primary to an
earlier type of O6.5, or even O6, depending on the contribution to the
He~\1 spectrum by the secondary; therefore we adopt the type employed
by \citeauthor{bmn99} of O6-7 V((f)).

For the remaining 7 stars studied by \citeauthor{bmn99}, our radial
velocities are in excellent agreement.  However, the comparison of the
H$\alpha$ Balmer line profiles (see Figure~\ref{halpha}) highlights
the disagreement between the FLAMES and ISIS spectra of 6611-011
(W314); no differences were revealed between the two FLAMES
observations.  Closer inspection of the blue-region FLAMES data
reveals slightly peculiar profiles for a number of the He~\1 lines,
namely \lam\lam4026, 4121, 4388 and 4471, with possibly weak
line-doubling seen in the latter two.  Both Trumpler
\citep[from][]{w61} and \citeauthor{bmn99} also suggested possible secondary components.
The exact nature of this system is still not entirely clear but the
new data here, in particular the H$\alpha$ profiles in
Figure~\ref{halpha}, provide further evidence of a possible companion.
That the lines are seen to vary slightly between different epochs may
also account for the slightly earlier classification assigned here to
the FLAMES-Giraffe spectrum of O9~V \citep[cf. B0~V
from][]{hmsm,bmn99}.

We have only one observational epoch for NGC\,3293 and NGC\,4755,
but inspection of the FLAMES spectra revealed a double-lined
binary, namely 4755-024.  The principal component appears as a
relatively broad-lined B2.5 dwarf, whilst the relative ratio of the
He~\1 \lam4471 line to the Mg~\2 \lam4481 in the secondary spectrum
suggests a spectral type of approximately B5 for the companion.

In comparison to published values, the radial velocities in Section
\ref{rvs} also provide an opportunity to detect potential binaries in
NGC\,3293 and NGC\,4755.  \citet{f3293} published velocities for a
number of stars in NGC\,3293, 20 of which were observed with FEROS or
FLAMES.  In general there is reasonable agreement between the two sets
of results, but a few stars have significantly discrepant ($|\Delta
v_{\rm r}| >$ 20 \kms) velocities.  In NGC\,3293 these stars are
3293-008 (F007), 3293-013 (F005), and 3293-014 (F019), all of which
were identified by Feast as having variable velocities.  Furthermore,
\citeauthor{f3293} suggested that 3293-029 (F028) displayed evidence of
line doubling and that it was probably a binary.  The FLAMES-Giraffe
spectrum of this star displays broad diffuse lines, with some very
mild asymmetry in the weaker He~\1 lines such as \lam4144 and
\lam4388.  Its radial velocity (8~\kms) is different to Feast's
singular measurement ($-$13~\kms), but his value was flagged as
uncertain and the true nature of this star is still not clear.

In NGC\,4755 our results are in good agreement with those of
\citet{her60}.  Comparison with the results from \citet{f4755} again
highlights a number of potential binaries, namely, 4755-005 (II-23),
4755-007 (III-05), 4755-012 (III-07), 4755-013 (IV-18) and 4755-021
(I), where the aliases are again taken from \citet{as58}.
Interestingly \citeauthor{f4755} presented many measurements for
4755-005, commenting that it is a double-lined binary; our derived
velocity of $-$55~\kms\/ is different to those tabulated by Feast, and
the spectrum does not appear to contain a secondary component.

\section{Summary}
We have introduced our FLAMES survey of early-type stars in Galactic
and Magellanic Cloud open clusters, in particular explaining the
philosophy employed in target selection in the Galactic fields.  For
one of the clusters, NGC\,3293, we have reduced the FLAMES data using
the pipeline reduction software (girBLDRS) and also using more
manual methods in IRAF, finding excellent agreement in the final
rectified spectra.  Similarly, longslit ISIS spectra of stars in
NGC\,6611 are also in excellent agreement with the FLAMES data,
providing further confidence in the multi-fibre observations.

The spectra were classified, with discussion of specific
peculiarities, potential Be-type stars and evidence of binarity.  In
particular, with the benefit of the high-resolution FLAMES spectra we
have detected interstellar He~\1 toward some of the NGC\,6611
stars, indicative of some highly ionized plasma in the region of the
cluster.  Stellar radial velocities are given for the majority of our
targets, which help to highlight cluster non-members and further
potential binaries.

The distributions of spectral types for the three Galactic clusters
are summarized in Table \ref{sp_sum}; in total we have observed 319
Galactic stars.  These observations represent a significant increase
in the known spectral content of NGC\,3293 and NGC\,4755, and will
serve as comparison stars for our forthcoming data in the Magellanic
Clouds.

\section{Acknowledgements}
We are extremely grateful to Francesca Primas and the staff at both
Paranal and La Silla for their invaluable assistance with the
observational programme, and to Andre Blecha and Gilles Simond at the
Observatoire de Gen$\grave{\rm e}$ve for their helpful responses to
our girBLDRS queries.  We also thank Nolan Walborn for numerous
insightful discussions, Damian Christian and Peter van Hoof for their
input on the identification of the non-stellar helium absorption, 
Cameron Reed for his input on the correct identification of CPD$-$59$^\circ$4553,
and the referee for their constructive comments.  CJE acknowledges
financial support from the UK Particle Physics and Astronomy Research
Council (PPARC) under grant PPA/G/S/2001/00131.  AH, SSD, and MRV
thank the Spanish Ministerio de Educacion y Ciencia under project
PNAYA 2001-0436.  FN acknowledges support through grants
AYA2003-02785E and ESP2002-01627.  This paper is based in part on
observations made with the Isaac Newton and William Herschel
telescopes, operated on the island of La Palma by the Isaac Newton
Group in the Spanish Observatorio del Roque de los Muchachos of the
Instituto de Astrof\'{\i}sica de Canarias.

{\scriptsize
\begin{center}
\begin{longtable}{llcccclrll}
\caption[]{NGC3293: Observational Parameters of Target Stars.  
The identifications in column two are from: \citet[F, ][]{f3293};
\citet[T, ][]{turn80}; \citet[HM, ][]{hm82}.
WFI photometry is quoted for $V >$ 10.75$^{\rm m}$, brighter than this
there were saturation problems and values are from \citeauthor{turn80}
and \citet{fm90}.  Radial velocities ($v_{\rm r}$) are given in \kms.
Instrument codes refer to FEROS (F) and Giraffe (G).
\label{3293}} \\
\hline
ID & Alias & $\alpha$(2000) & $\delta$(2000) & $V$ & $B-V$ & Sp. Type & $v_{\rm r}$  & Inst. & Comments \\
\hline 
\endfirsthead
\caption[]{\it{continued}} \\
\hline
ID & Alias & $\alpha$(2000) & $\delta$(2000) & $V$ & $B-V$ & Sp. Type & $v_{\rm r}$  & Inst & Comments \\
\hline 
\endhead
\hline 
\multicolumn{10}{r}{\it{continued on next page}} \\
\endfoot
\hline 
\endlastfoot
\hline
3293-001 & F004 &  10 35 49.33  & $-$58 13 27.6  &\o6.52  & 0.00   & B0 Iab          &     0\p &  F    & HD~91969, CPD$-$57$^\circ$3508; H$\alpha$ =  abs$+$em.(stellar?)\\
3293-002 & F003 &  10 35 42.02  & $-$58 11 34.8  &\o6.73  & 0.07   & B0.7 Ib         &  $-$7\p &  F     & HD~91943, CPD$-$57$^\circ$3499\\
3293-003 & F022 &  10 35 46.57  & $-$58 14 11.7  &\o7.61  & 0.11   & B1 III          & $-$16\p &  F     & CPD$-$57$^\circ$3506A\\
3293-004 & F020 &  10 35 57.72  & $-$58 13 20.6  &\o8.03  & 0.02   & B1 III          & $-$21\p &  F     & CPD$-$57$^\circ$3523\\ 
3293-005 & F025 &  10 35 56.55  & $-$58 14 34.7  &\o8.12  & 0.08   & B1 III          &  $-$8\p &  F     & CPD$-$57$^\circ$3521\\ 
3293-006 & F006 &  10 35 58.88  & $-$58 14 26.1  &\o8.21  & 0.07   & B1 III          & $-$10\p &  F     & CPD$-$57$^\circ$3526B\\
3293-007 & F008 &  10 36 16.08  & $-$58 16 37.9  &\o8.25  & 0.18   & B1 III          & $-$16\p &  F     & HD~92044, CPD$-$57$^\circ$3540\\ 
3293-008 & F007 &  10 35 54.22  & $-$58 15 26.7  &\o8.59  & 0.05   & B1 III          &    57\p &  F     & HD~91983, CPD$-$57$^\circ$3516\\ 
3293-009 & F048 &  10 36 46.90  & $-$58 17 53.3  &\o8.64  & 0.23   & A7 III          &  $-$5\p &  F     & HD~92121, CPD$-$57$^\circ$3563\\
3293-010 & F016 &  10 35 40.72  & $-$58 12 44.0  &\o8.77  & 0.00   & B1 III          & $-$16\p &  F     & CPD$-$57$^\circ$3500\\ 
3293-011 & F026 &  10 35 58.49  & $-$58 14 14.8  &\o8.85  & 0.04   & Be (B1:)        &     $-$ &  F     & CPD$-$57$^\circ$3526; H$\alpha$ = broad em.\\ 
3293-012 & F027 &  10 36 01.60  & $-$58 15 09.4  &\o8.95  & 0.06   & B1 III          & $-$28\p &  F     & HD~92007, CPD$-$57$^\circ$3527\\
3293-013 & F005 &  10 36 08.34  & $-$58 13 04.1  &\o9.03&$-$0.04\pp& B1 III          & $-$60\p &  F     & HD~92024, CPD$-$57$^\circ$3533\\ 
3293-014 & F019 &  10 35 58.63  & $-$58 12 32.1  &\o9.09  & 0.19   & B0.5 IIIn       &    50\p &  F     & CPD$-$57$^\circ$3524A\\ 
3293-015 & T133 &  10 35 54.92  & $-$58 12 58.9  &\o9.11&$-$0.01\pp& B1 V            & $-$16\p &  F     & CPD$-$57$^\circ$3515\\ 
3293-016 & F023 &  10 35 48.45  & $-$58 14 14.5  &\o9.21  & 0.03   & B2.5 V          & $-$14\p &  G     & CPD$-$57$^\circ$3506B; strongly blended\\
3293-017 & F024 &  10 35 53.65  & $-$58 14 47.6  &\o9.22  & 0.04   & B1 V            &  $-$4\p &  F     & CPD$-$57$^\circ$3517\\
3293-018 & F018 &  10 35 57.81  & $-$58 12 21.1  &\o9.26&$-$0.04\pp& B1 V            &  $-$7\p &  F     & CPD$-$57$^\circ$3524B\\
3293-019 & F014 &  10 35 48.22  & $-$58 12 32.9  &\o9.27&$-$0.04\pp& B1 V            & $-$17\p &  F     & CPD$-$57$^\circ$3507\\ 
3293-020 & F010 &  10 35 30.07  & $-$58 12 08.0  &\o9.55  &  0.03  & B1.5 III        & $-$16\p &  G     & HDE~303067\\
3293-021 & T065 &  10 35 43.31  & $-$58 13 33.4  &\o9.85  &  0.03  & B1.5 III        &  $-$8\p &  G     & CPD$-$57$^\circ$3503\\
3293-022 & F012 &  10 35 32.30  & $-$58 15 22.0  &\o9.97  &  0.12  & Be (B0.5-1.5n)  & $-$20\p &  G     & HDE~303075; H$\alpha$ = twin\\
3293-023 & T131 &  10 35 55.39  & $-$58 12 19.7  & 10.01&$-$0.05\pp& B1.5 III        & $-$10\p &  G     & CPD$-$57$^\circ$3519\\
3293-024 & F009 &  10 36 04.91  & $-$58 10 43.3  & 10.01&$-$0.01\pp& B1.5 III        & $-$14\p &  G     & HDE~303065\\
3293-025 & F015 &  10 35 45.17  & $-$58 12 23.6  & 10.01  &  0.00  & B2 III          & $-$18:  &  G     & CPD$-$57$^\circ$3504; strongly blended\\
3293-026 & F013 &  10 35 56.60  & $-$58 11 31.4  & 10.16&$-$0.02\pp& B2 III          & $-$21\p &  G     & CPD$-$57$^\circ$3522\\
3293-027 & T079 &  10 35 39.97  & $-$58 13 56.9  & 10.22  &  0.07  & Be (B0.5-1.5n)  & $-$13\p &  G     & H$\alpha$ = abs$+$twin\\
3293-028 & T132 &  10 35 56.61  & $-$58 12 40.7  & 10.26  &  0.01  & B2 V            & $-$10\p &  G     & CPD$-$57$^\circ$3520\\
3293-029 & F028 &  10 36 05.95  & $-$58 14 27.0  & 10.32&$-$0.01\pp& B0.5-B1.5 Vn    &     8\p &  G     & CPD$-$57$^\circ$3531\\
3293-030 & F017 &  10 35 53.01  & $-$58 12 16.8  & 10.51&$-$0.04\pp& B2 V            &  $-$7\p &  G     & CPD$-$57$^\circ$3514\\
3293-031 & F033 &  10 36 03.51  & $-$58 14 40.0  & 10.66  &  0.05  & B2 V            &  $-$7\p &  G     & CPD$-$57$^\circ$3528\\ 
3293-032 & F029 &  10 35 54.67  & $-$58 13 48.6  & 10.69  &  0.01  & B0.5-B1.5 Vn    &    0\p  &  G     & CPD$-$57$^\circ$3518\\
3293-033 & F041 &  10 34 33.88  & $-$58 12 28.2  & 10.72  &  0.66  & B8 III          &     8\p &  G     & HDE~303073\\
3293-034 & T086 &  10 35 49.01  & $-$58 14 54.1  & 10.74  &  0.11  & B2 IIIh         & $-$16\p &  G     & \\
3293-035 & F045 &  10 36 40.29  & $-$58 05 54.7  & 10.81  &  0.09  & B2 V            &  $-$9\p &  G     & HDE~303062\\
3293-036 & F036 &  10 36 17.02  & $-$58 08 26.5  & 10.90  &  0.19  & A5 III          &    29:  &  G     & HDE~303066\\
3293-037 & T089 &  10 36 07.64  & $-$58 15 20.2  & 10.94  &  0.04  & B2 V            & $-$12:  &  G     & \\
3293-038 & $-$  &  10 35 06.64  & $-$58 10 34.7  & 11.00&$-$0.04\pp& B2.5 V          & $-$13:  &  G     & \\
3293-039 & F051 &  10 37 04.06  & $-$58 08 01.3  & 11.07  &  0.07  & A3 III          &    10\p &  G     & HDE~303063\\
3293-040 & T063 &  10 35 44.05  & $-$58 13 45.9  & 11.21  &  0.05  & Be (B3n)        & $-$12\p &  G     & H$\alpha$ = abs$+$twin\\
3293-041 & T087 &  10 35 44.66  & $-$58 14 30.1  & 11.22  &  0.04  & B2.5 V          & $-$12\p &  G     & \\
3293-042 & F044 &  10 35 58.93  & $-$58 05 17.8  & 11.27  &  0.11  & A3 III          &  $-$5\p &  G     & HDE~303061\\
3293-043 & T092 &  10 35 28.51  & $-$58 12 49.6  & 11.32  &  0.07  & B3 V            &     0\p &  G     & \\
3293-044 & F039 &  10 34 38.31  & $-$58 07 42.3  & 11.35  &  0.10  & A3 III          &  $-$9:  &  G     & HDE~303071\\
3293-045 & $-$  &  10 36 13.70  & $-$58 17 32.7  & 11.42  &  0.32  & Be (B1-2n)      & $-$18\p &  G     & H$\alpha$ = twin\\
3293-046 & F046 &  10 35 15.05  & $-$58 04 26.8  & 11.44  &  0.11  & A7 III          &     9\p &  G     & HDE~303060\\
3293-047 & HM322&  10 35 36.62  & $-$58 16 04.0  & 11.55  &  0.13  & B2.5 V          & $-$16\p &  G     & \\
3293-048 & HM292&  10 35 44.99  & $-$58 16 35.0  & 11.56  &  0.16  & B2.5 V          & $-$12\p &  G     & \\
3293-049 & $-$  &  10 34 20.79  & $-$58 13 30.5  & 11.64  &  0.07  & B2.5 V          & $-$18\p &  G     & \\
3293-050 & T075 &  10 35 41.94  & $-$58 11 56.7  & 11.69  &  0.05  & B3 Vn           & $-$12\p &  G     & \\
3293-051 & $-$  &  10 34 38.85  & $-$58 07 29.7  & 11.73  &  0.20  & A5 III          &  $-$1\p &  G     & \\
3293-052 & HM459&  10 35 29.91  & $-$58 09 28.3  & 11.77  &  0.22  & A3 III          &  $-$3\p &  G     & \\
3293-053 & T130 &  10 35 50.74  & $-$58 11 41.2  & 11.83  &  0.03  & B3 V            & $-$15\p &  G     & \\
3293-054 & $-$  &  10 36 25.86  & $-$58 14 36.2  & 11.85  &  0.06  & A0 III          & $-$13\p &  G     & \\
3293-055 & HM212&  10 36 28.66  & $-$58 13 35.5  & 11.88  &  0.12  & A3 III          &     1\p &  G     & \\
3293-056 & $-$  &  10 35 13.55  & $-$58 11 12.4  & 11.91  &  0.00  & B3 V            &  $-$2\p &  G     & \\
3293-057 & T084 &  10 35 45.30  & $-$58 15 28.0  & 11.92  &  0.21  & B3 V            & $-$11\p &  G     & faint companion in WFI image\\
3293-058 & $-$  &  10 35 38.86  & $-$58 04 26.3  & 11.99  &  0.08  & A0 IIp (Si)     &    12\p &  G     & \\
3293-059 & HM277&  10 36 06.57  & $-$58 17 53.8  & 12.00  &  0.29  & B5 III-Vn       &     0\p &  G     & \\
3293-060 & HM231&  10 35 21.10  & $-$58 12 00.5  & 12.02  &  0.07  & A2 III          &     8\p &  G     & \\
3293-061 & T052 &  10 35 54.72  & $-$58 12 37.0  & 12.03  &  0.05  & B5 V            & $-$14\p &  G     & \\
3293-062 & HM264&  10 36 05.25  & $-$58 16 45.5  & 12.03  &  0.15  & B3 V            & $-$62\p &  G     & \\
3293-063 & $-$  &  10 37 04.82  & $-$58 10 08.1  & 12.05  &  0.09  & B5 V            & $-$10\p &  G     & \\
3293-064 & $-$  &  10 35 54.40  & $-$58 12 56.3  & 12.06  &  0.06  & A0 II           &     3\p &  G     & adjacent to T133 \\
3293-065 & $-$  &  10 34 42.02  & $-$58 15 41.9  & 12.06  &  0.14  & B5 III-V        & $-$18\p &  G     & \\ 
3293-066 & T090 &  10 35 57.12  & $-$58 15 21.6  & 12.07  &  0.16  & B5 V            & $-$49:  &  G     & \\
3293-067 & T117 &  10 36 13.48  & $-$58 11 20.7  & 12.11  &  0.32  & B3 V            & $-$18:  &  G     & \\
3293-068 & $-$  &  10 35 09.57  & $-$58 07 47.7  & 12.21  &  0.06  & B9 III          &    12\p &  G     & \\
3293-069 & HM393&  10 35 24.10  & $-$58 13 28.4  & 12.22  &  0.07  & B5 V            & $-$16\p &  G     & \\
3293-070 & $-$  &  10 34 54.80  & $-$58 07 23.3  & 12.22  &  0.14  & B5 III-V        &    12\p &  G     & \\
3293-071 & $-$  &  10 34 48.62  & $-$58 13 13.4  & 12.23  &  0.22  & A7 II-III       & $-$14\p &  G     & \\
3293-072 & HM408&  10 35 16.11  & $-$58 12 59.7  & 12.25  &  0.00  & B8 IIp          & $-$14\p &  G     & composite spectrum?\\
3293-073 & $-$  &  10 35 38.51  & $-$58 13 06.8  & 12.25  &  0.11  & B6-7 V          & $-$16\p &  G     & \\
3293-074 & HM403&  10 35 21.96  & $-$58 12 53.6  & 12.28  &  0.09  & B8 III          & $-$16:  &  G     & \\
3293-075 & HM299&  10 35 22.72  & $-$58 17 08.1  & 12.29  &  0.12  & B5 IIIn         & $-$11\p &  G     & \\
3293-076 & $-$  &  10 35 02.75  & $-$58 13 03.4  & 12.29  &  0.34  & F0 III          &     7\p &  G     & \\
3293-077 & $-$  &  10 34 58.30  & $-$58 11 18.6  & 12.32  &  0.20  & B6-7 V          &  $-$1\p &  G     & no \lam4656, 6515 data\\
3293-078 & $-$  &  10 36 07.04  & $-$58 05 25.6  & 12.36  &  0.13  & A3 III          &  $-$9\p &  G     & \\
3293-079 & HM352&  10 35 23.68  & $-$58 15 06.1  & 12.36  &  0.13  & A0 III          &  $-$2\p &  G     & \\
3293-080 & $-$  &  10 36 06.56  & $-$58 06 55.1  & 12.36  &  0.14  & B5 V            & $-$20\p &  G     & \\
3293-081 & $-$  &  10 36 45.39  & $-$58 09 50.5  & 12.38  &  0.32  & A5 II           & $-$24:  &  G     & \\
3293-082 & $-$  &  10 35 47.53  & $-$58 05 13.3  & 12.40  &  0.21  & B5 III          &  $-$8:  &  G     & \\
3293-083 & $-$  &  10 36 33.50  & $-$58 20 33.8  & 12.46  &  0.28  & A3 III          &     6:  &  G     & \\
3293-084 & T093 &  10 36 02.29  & $-$58 12 58.1  & 12.48  &  0.11  & B5 V            & $-$17\p &  G     & \\
3293-085 & $-$  &  10 36 28.00  & $-$58 08 18.0  & 12.49  &  0.13  & B5 V            & $-$10\p &  G     & \\
3293-086 & $-$  &  10 35 54.92  & $-$58 06 49.3  & 12.52  &  0.16  & B5 V            & $-$13:  &  G     & \\
3293-087 & $-$  &  10 35 04.74  & $-$58 14 15.4  & 12.60  &  0.04  & B5 V            & $-$14\p &  G     & \\
3293-088 & HM401&  10 35 21.74  & $-$58 12 37.8  & 12.60  &  0.24  & F0 III-V        &     5\p &  G     & \\
3293-089 & $-$  &  10 36 50.05  & $-$58 04 52.3  & 12.64  &  0.15  & B8 III          & $-$13\p &  G     & \\
3293-090 & T104 &  10 35 45.81  & $-$58 16 00.0  & 12.67  &  0.31  & B6-7 V          & $-$11\p &  G     & \\
3293-091 & $-$  &  10 35 59.03  & $-$58 05 53.7  & 12.68  &  0.34  & A7 II           & $-$20\p &  G     & \\
3293-092 & $-$  &  10 36 15.02  & $-$58 08 04.5  & 12.70  &  0.16  & B9 III          &  $-$9\p &  G     & \\
3293-093 & T081 &  10 35 35.68  & $-$58 13 56.4  & 12.71  &  0.14  & B6-7 V          & $-$12\p &  G     & \\
3293-094 & T121 &  10 36 17.91  & $-$58 14 29.5  & 12.71  &  0.20  & B5 V            & $-$34:  &  G     & \\
3293-095 & T120 &  10 36 12.90  & $-$58 13 24.9  & 12.72  &  0.13  & B6-7 V          & $-$18\p &  G     & \\
3293-096 & $-$  &  10 35 03.25  & $-$58 14 26.8  & 12.74  &  0.23  & B6-7 III        &    12:  &  G     & \\
3293-097 & T124 &  10 36 16.92  & $-$58 15 00.0  & 12.75  &  0.13  & B6-7 III        & $-$10\p &  G     & faint companion in WFI image\\
3293-098 & T055 &  10 35 52.82  & $-$58 13 11.7  & 12.75  &  0.14  & B8 III-V        & $-$13\p &  G     & \\
3293-099 & $-$  &  10 36 46.36  & $-$58 09 29.1  & 12.78  &  0.18  & B5 Vn           & $-$14\p &  G     & \\
3293-100 & HM313&  10 35 51.31  & $-$58 16 17.8  & 12.81  &  0.14  & B6-7 III-Vn     & $-$15:  &  G     & \\
3293-101 & $-$  &  10 36 47.50  & $-$58 18 01.2  & 12.82  &  0.36  & F0 III          &    16\p &  G     & \\
3293-102 & $-$  &  10 34 53.56  & $-$58 20 58.7  & 12.83  &  0.28  & A2 III          &    16\p &  G     & \\
3293-103 & $-$  &  10 34 57.00  & $-$58 08 06.0  & 12.83  &  0.29  & A7 II           &     2\p &  G     & \\
3293-104 & T069 &  10 35 40.29  & $-$58 13 01.0  & 12.88  &  0.11  & B6-7 V          &  $-$8:  &  G     & \\
3293-105 & $-$  &  10 35 57.85  & $-$58 15 02.8  & 12.89  &  0.19  & B8 III-V        & $-$11\p &  G     & \\
3293-106 & F034 &  10 35 50.51  & $-$58 12 12.6  & 12.90  &  0.08  & B6-7 V          & $-$22\p &  G     & \\
3293-107 & HM243&  10 36 10.82  & $-$58 15 48.9  & 12.91  &  0.22  & B8 III-V        & $-$11\p &  G     & \\
3293-108 & $-$  &  10 36 09.85  & $-$58 05 44.3  & 12.92  &  0.13  & B6-7 V          &  $-$6\p &  G     & \\
3293-109 & HM454&  10 35 32.89  & $-$58 10 16.2  & 12.92  &  0.15  & B5 V            & $-$11\p &  G     & \\
3293-110 & $-$  &  10 37 02.26  & $-$58 12 37.1  & 12.93  &  0.28  & A3 II           & $-$19\p &  G     & \\
3293-111 & T091 &  10 35 55.41  & $-$58 15 37.1  & 12.96  &  0.16  & B6-7 III-V      & $-$13\p &  G     & \\
3293-112 & $-$  &  10 34 34.76  & $-$58 16 05.0  & 12.99  &  0.18  & A5 III          &    13:  &  G     & \\
3293-113 & T100 &  10 35 58.37  & $-$58 14 40.9  & 13.04  &  0.14  & B6-7 V          & $-$16\p &  G     & \\
3293-114 & $-$  &  10 36 38.81  & $-$58 13 57.4  & 13.07  &  0.25  & B6-7 III-V      & $-$15\p &  G     & \\
3293-115 & $-$  &  10 35 53.98  & $-$58 15 04.4  & 13.11  &  0.14  & B8 III-V        & $-$13\p &  G     & \\
3293-116 & T071 &  10 35 35.84  & $-$58 13 07.6  & 13.12  &  0.11  & B6-7 V          & $-$14\p &  G     & \\
3293-117 & T114 &  10 36 09.76  & $-$58 10 57.8  & 13.15  &  0.14  & B8 III-V        &  $-$8\p &  G     & \\
3293-118 & $-$  &  10 35 06.71  & $-$58 05 07.7  & 13.17  &  0.06  & B8 III-V        & $-$10\p &  G     & \\
3293-119 & T109 &  10 35 36.70  & $-$58 15 28.2  & 13.18  &  0.22  & B9 III          & $-$12\p &  G     & \\
3293-120 & $-$  &  10 34 47.74  & $-$58 07 27.4  & 13.20  &  0.33  & B5 V            & $-$12:  &  G     & \\
3293-121 & HM316&  10 35 41.32  & $-$58 15 39.2  & 13.23  &  0.21  & B8: III         &  $-$9\p &  G     & \\
3293-122 & T119 &  10 36 18.64  & $-$58 12 53.5  & 13.24  &  0.15  & B9 III          & $-$12\p &  G     & \\
3293-123 & $-$  &  10 35 53.12  & $-$58 14 26.0  & 13.26  &  0.20  & B8 III          & $-$27\p &  G     & \\
3293-124 & $-$  &  10 35 44.93  & $-$58 05 11.2  & 13.28  &  0.21  & B8 III          &  $-$9\p &  G     & \\
3293-125 & T067 &  10 35 47.53  & $-$58 12 46.9  & 13.29  &  0.10  & B8 III-V        & $-$14:  &  G     & \\
3293-126 & HM285&  10 35 30.63  & $-$58 17 40.4  & 13.29  &  0.22  & A7 III          &     0\p &  G     & \\
\hline

\end{longtable}
\end{center}
}

{\scriptsize
\begin{center}
\begin{longtable}{llcccclrll}
\caption[]{NGC4755: Observational Parameters of Target Stars.  
The identifications in column two are those of \citet{as58}; uncertain
matches from the finding charts are denoted by `$\ast$'.  WFI
photometry is quoted for $V >$ 11.5$^{\rm m}$, brighter than this there were
saturation problems and values are from \citet{sb01}.  Radial
velocities ($v_{\rm r}$) are given in \kms.  Instrument codes refer to
FEROS (F) and Giraffe (G).
\label{4755}} \\
\hline
ID & Alias & $\alpha$(2000) & $\delta$(2000) & $V$ & $B-V$ & Sp. Type & $v_{\rm r}$  & Inst. & Comments \\
\hline 
\endfirsthead
\caption[]{\it{continued}} \\
\hline
ID & Alias & $\alpha$(2000) & $\delta$(2000) & $V$ & $B-V$ & Sp. Type & $v_{\rm r}$  & Inst. & Comments \\
\hline 
\endhead
\hline 
\multicolumn{10}{r}{\it{continued on next page}} \\
\endfoot
\hline 
\endlastfoot
\hline
4755-001 & A                 &  12 53 21.92 & $-$60 19 47.4 &\o5.77 &  0.32   & B9 Ia        & $-$16\p & F     & HD\,111904, CPD$-$59$^\circ$4529\\
4755-002 & B  &  12 53 48.94 & $-$60 22 36.8 &\o5.98 &  0.22   & B3 Ia        & $-$10\p &  F& $\kappa$ Crucis; HD\,111973, CPD$-$59$^\circ$4555; H$\alpha$ = broad em.\\ 
4755-003 & C                 &  12 53 59.81 & $-$60 20  8.7 &\o6.80 &  0.24   & B2 III       & $-$21\p &  F  & HD\,111990, CPD$-$59$^\circ$4566; H$\alpha$=abs$+$em (stellar?)\\
4755-004 & I$-$06            &  12 53 37.64 & $-$60 21 25.7 &\o6.92 &  0.20   & B1.5 Ib      & $-$19\p &  F     & HD\,111934, CPD$-$59$^\circ$4543\\
4755-005 & II$-$23           &  12 53 47.30 & $-$60 19 55.6 &\o7.96 &  0.20   & B2 III       & $-$53\p &  F     & CPD$-$59$^\circ$4551\\
4755-006 & E                 &  12 53 46.52 & $-$60 24 12.6 &\o8.37 &  0.14   & B1 III       & $-$20\p &  F     & CPD$-$59$^\circ$4552\\ 
4755-007 & III$-$05          &  12 53 49.52 & $-$60 23  3.1 &\o8.58 &  0.11   & B1 V         & $-$12\p &  F     & CPD$-$59$^\circ$4557\\
4755-008 & I$-$05            &  12 53 39.20 & $-$60 21 13.0 &\o8.66 &  0.14   & B0.5 V       &  $-$2\p &  F     & \\ 
4755-009 & F                 &  12 53 57.58 & $-$60 24 58.3 &\o9.01 &  0.11   & B1 V         & $-$15\p &  F     & CPD$-$59$^\circ$4564\\ 
4755-010 & II$-$01           &  12 53 52.03 & $-$60 22 16.0 &\o9.38 &  0.16   & B1 V         &     0\p &  F     & \\
4755-011 & R                 &  12 53 46.57 & $-$60 22 18.5 &\o9.58 &  0.17   & B1.5 V       & $-$24\p &  F     & \\
4755-012 & III$-$07          &  12 53 52.27 & $-$60 22 28.0 &\o9.60 &  0.16   & B1.5 V       & $-$55\p &  F     & CPD$-$59$^\circ$4560\\ 
4755-013 & IV$-$18           &  12 53 35.52 & $-$60 23 46.9 &\o9.68 &  0.15   & B1.5 V       &  $-$3\p &  F     & CPD$-$59$^\circ$4542\\
4755-014 & S                 &  12 53 47.28 & $-$60 22 20.2 &\o9.72 &  0.20   & Be (B1:)     &   $-$\p &  F     & H$\alpha$ = twin\\
4755-015 & G                 &  12 53 20.69 & $-$60 23 17.1 &\o9.74 &  0.16   & B1 V         & $-$18\p &  F     & CPD$-$59$^\circ$4528\\ 
4755-016 & III$-$01          &  12 53 43.91 & $-$60 22 29.5 &\o9.76 &  0.12   & B1.5 V       &  $-$8\p &  F     & CPD$-$59$^\circ$4549\\
4755-017 & H                 &  12 53 33.27 & $-$60 24 33.5 &\o9.90 &  0.18   & B1.5 V       & $-$15\p &  F     & CPD$-$59$^\circ$4540\\
4755-018 & IV$-$17           &  12 53 39.02 & $-$60 23 43.8 &\o9.96 &  0.16   & Be (B1.5:)   &   $-$\p &  F     & CPD$-$59$^\circ$4546; H$\alpha$ = broad em.\\
4755-019 & IV$-$52           &  12 53 10.23 & $-$60 25 59.8 & 10.04 &  0.17   & B1.5 V       & $-$20\p &  F     & HDE 312079, CPD$-$59$^\circ$4523\\ 
4755-020 & IV$-$54           &  12 53 14.25 & $-$60 27 38.9 & 10.05 &  0.13   & B2 V         & $-$27\p &  G     & \\
4755-021 & I                 &  12 53 47.02 & $-$60 18 35.7 & 10.29 &  0.20   & B1.5 Vn      & $-$11\p &  F     & CPD$-$59$^\circ$4550\\
4755-022 & I$-$17            &  12 53 24.28 & $-$60 21 31.0 & 10.89 &  0.12   & B2.5 Vn      & $-$18\p &  G     & CPD$-$59$^\circ$4531\\
4755-023 & J                 &  12 53 25.96 & $-$60 20 48.1 & 10.99 &  0.18   & B2 V         & $-$26\p &  G     & CPD$-$59$^\circ$4537\\
4755-024 & III$-$32          &  12 53 56.66 & $-$60 26 32.9 & 11.00 &  0.12   & B2.5 V + B5: & $-$85\p &  G     & CPD$-$59$^\circ$4563\\
4755-025 & III$-$26          &  12 54 03.63 & $-$60 25 21.1 & 11.02 &  0.14   & B2.5 V       & $-$20\p &  G     & \\
4755-026 & IV$-$03           &  12 53 39.41 & $-$60 22 40.0 & 11.19 &  0.24   & B2.5 V       & $-$15\p &  G     & \\
4755-027 & II$-$14           &  12 53 43.22 & $-$60 20 47.5 & 11.21 &  0.17   & B2.5 Vn      & $-$18\p &  G     & \\
4755-028 & I$-$50            &  12 53 25.39 & $-$60 16 18.3 & 11.22 &  0.31   & B9 III       &     1\p &  G     & \\
4755-029 & I$-$38            &  12 53 25.43 & $-$60 19 11.5 & 11.23 &  0.19   & B2 V         & $-$19\p &  G     & CPD$-$59$^\circ$4536\\
4755-030 & II$-$20           &  12 54 08.61 & $-$60 21 38.7 & 11.31 &  0.21   & B2.5 Vn      & $-$35\p &  G     & \\
4755-031 & IV$-$59           &  12 52 50.69 & $-$60 25 36.9 & 11.34 &  0.24   & B2.5 V       & $-$20\p &  G     & \\
4755-032 & IV$-$26           &  12 53 09.90 & $-$60 22 27.5 & 11.35 &  0.19   & B2.5 V       & $-$13\p &  G     & \\
4755-033 & IV$-$11           &  12 53 39.95 & $-$60 23 27.1 & 11.35 &  0.24   & B3 V         & $-$19\p &  G     & \\
4755-034 & K                 &  12 53 22.61 & $-$60 23 47.8 & 11.39 &  0.23   & B3 V         & $-$19\p &  G     & CPD$-$59$^\circ$4530\\
4755-035 & I$-$07            &  12 53 38.21 & $-$60 21 45.3 & 11.41 &  0.23   & B5 V         & $-$31\p &  G     & \\
4755-036 & III$-$16          &  12 53 43.36 & $-$60 24 02.5 & 11.45 &  0.16   & B3 Vn        & $-$29\p &  G     & \\
4755-037 & III$-$18          &  12 53 47.01 & $-$60 25 18.4 & 11.48 &  0.13   & B2.5 V       & $-$19\p &  G     & \\
4755-038 & III$-$13          &  12 53 51.83 & $-$60 23 54.5 & 11.54 &  0.16   & B3 Ve        & $-$19\p &  G     & H$\alpha$ = v. weak, twin\\
4755-039 & IV$-$06           &  12 53 37.13 & $-$60 22 55.4 & 11.54 &  0.16   & B2.5 V       & $-$23\p &  G     & \\
4755-040 & $-$               &  12 54 22.49 & $-$60 13 14.4 & 11.54 &  0.32   & B2.5 V       & $-$20\p &  G     & \\
4755-041 & II$-$13           &  12 53 44.05 & $-$60 20 58.1 & 11.58 &  0.21   & B3 V         & $-$16\p &  G     & \\
4755-042 & I$-$39            &  12 53 26.48 & $-$60 19 00.5 & 11.58 &  0.29   & B3 Vn        & $-$14\p &  G     & \\
4755-043 & II$-$15           &  12 53 49.41 & $-$60 20 57.6 & 11.59 &  0.22   & B3 V         & $-$29\p &  G     & \\
4755-044 & $-$               &  12 54 44.34 & $-$60 25 14.2 & 11.59 &  0.28   & B5 V         & $-$12\p &  G     & \\
4755-045 & IV$-$08           &  12 53 32.47 & $-$60 22 39.3 & 11.61 &  0.22   & B3 V         & $-$17\p &  G     & \\
4755-046 & $-$               &  12 54 07.16 & $-$60 23 13.4 & 11.62 &  0.20   & B3 Vn        & $-$29:  &  G     & \\
4755-047 & IV$-$13           &  12 53 26.20 & $-$60 22 52.2 & 11.62 &  0.23   & B3 V         & $-$20\p &  G     & \\
4755-048 & IV$-$47           &  12 53 09.66 & $-$60 24 13.7 & 11.63 &  0.19   & B3 V         & $-$20\p &  G     & \\
4755-049 & $-$               &  12 52 54.70 & $-$60 15 16.4 & 11.65 &  0.34   & B5 V         & $-$36:  &  G     & \\
4755-050 & IV$-$53           &  12 53 16.55 & $-$60 26 11.7 & 11.65 &  0.36   & A2 IIe       & $-$23\p &  G     & H$\alpha$ = abs$+$twin\\
4755-051 & $-$               &  12 53 53.09 & $-$60 23 07.4 & 11.75 &  0.23   & B3 Vn        & $-$28\p &  G     & \\
4755-052 & IV$-$42           &  12 53 33.93 & $-$60 26 22.4 & 11.77 &  0.17   & B3 V         & $-$20\p &  G     & \\
4755-053 & IV$-$27           &  12 53 41.83 & $-$60 24 38.4 & 11.78 &  0.18   & B3 Vn        & $-$22\p &  G     & \\
4755-054 & IV$-$46           &  12 53 14.03 & $-$60 24 13.4 & 11.78 &  0.19   & B3 V         & $-$20\p &  G     & \\
4755-055 & $-$               &  12 54 32.49 & $-$60 26 01.4 & 11.83 &  0.45   & A7 II        &  $-$8:  &  G     & \\
4755-056 & $-$               &  12 53 29.35 & $-$60 21 19.5 & 11.86 &  0.29   & B3 Vn        & $-$22\p &  G     & blended in WFI image\\
4755-057 & II$-$38           &  12 54 32.18 & $-$60 16 24.9 & 11.94 &  0.36   & B6-7 III-Ve  & $-$44:  &  G     & H$\alpha$ = abs$+$twin\\
4755-058 & II$-$07           &  12 53 45.46 & $-$60 21 08.3 & 12.07 &  0.26   & B3 Vn        & $-$28\p &  G     & \\
4755-059 & $-$               &  12 54 36.24 & $-$60 19 23.4 & 12.10 &  0.52   & F0 III       & $-$15\p &  G     & \\
4755-060 & IV$-$33           &  12 53 36.92 & $-$60 25 27.4 & 12.12 &  0.19   & B5 V         & $-$19\p &  G     & \\
4755-061 & I$-$22            &  12 53 08.94 & $-$60 21 22.7 & 12.15 &  0.25   & B5 V         & $-$20\p &  G     & strong Si~{\tiny II} for type\\
4755-062 & IV$-$31           &  12 53 17.92 & $-$60 23 26.3 & 12.18 &  0.28   & B6-7 III-V   & $-$14\p &  G     & \\
4755-063 & I$-$29            &  12 52 56.77 & $-$60 19 21.9 & 12.18 &  0.33   & B8 V         & $-$32\p &  G     & \\
4755-064 & I$-$09            &  12 53 32.24 & $-$60 22 19.7 & 12.22 &  0.27   & B6-7 V       & $-$23\p &  G     & \\
4755-065 & III$-$34          &  12 53 55.21 & $-$60 27 09.8 & 12.25 &  0.18   & B3 V         & $-$22\p &  G     & \\
4755-066 & $-$               &  12 52 14.46 & $-$60 25 49.3 & 12.25 &  0.27   & B3 V         & $-$24\p &  G     & \\
4755-067 & IV$-$30           &  12 53 33.47 & $-$60 24 20.7 & 12.28 &  0.24   & B5 V         & $-$23\p &  G     & \\ 
4755-068 & I$-$01            &  12 53 40.76 & $-$60 21 41.0 & 12.31 &  0.27   & B3 Vn        & $-$20\p &  G     & \\
4755-069 & I$-$36            &  12 53 25.35 & $-$60 20 21.3 & 12.31 &  0.28   & B5 V         & $-$17\p &  G     & \\
4755-070 & II$-$18           &  12 54 09.06 & $-$60 22 10.3 & 12.32 &  0.23   & B3 V         & $-$24\p &  G     & \\
4755-071 & $-$               &  12 53 33.78 & $-$60 20 54.3 & 12.35 &  0.32   & B9 III       & $-$24\p &  G     & \\
4755-072 & II$-$16           &  12 53 48.72 & $-$60 20 39.7 & 12.43 &  0.24   & B5 V         & $-$19\p &  G     & \\
4755-073 & I$-$16            &  12 53 36.13 & $-$60 20 32.5 & 12.45 &  0.24   & A0 II        & $-$15\p &  G     & \\
4755-074 & I$-$20            &  12 53 18.39 & $-$60 22 08.1 & 12.47 &  0.29   & B5 III-V     & $-$36:  &  G     & \\
4755-075 & I$-$11            &  12 53 25.87 & $-$60 22 14.9 & 12.52 &  0.24   & B5 III-V     & $-$22\p &  G     & \\
4755-076 & I$-$08            &  12 53 35.60 & $-$60 21 48.6 & 12.58 &  0.36   & B5 V         & $-$18\p &  G     & \\
4755-077 & I$-$37            &  12 53 37.51 & $-$60 19 26.8 & 12.59 &  0.32   & B6-7 III     & $-$27\p &  G     & \\
4755-078 & $-$               &  12 54 55.66 & $-$60 20 04.3 & 12.60 &  0.32   & B5 Vn        & $-$23\p &  G     & \\
4755-079 & I$-$18            &  12 53 19.54 & $-$60 21 31.0 & 12.62 &  0.28   & B5 V         & $-$20\p &  G     & \\
4755-080 & M                 &  12 53 18.19 & $-$60 18 49.8 & 12.62 &  0.35   & B8 III-V     & $-$19\p &  G     & \\
4755-081 & $-$               &  12 53 27.77 & $-$60 22 37.7 & 12.63 &  0.28   & B6-7 IIIn    & $-$18\p &  G     & \\
4755-082 & IV$-$25           &  12 53 17.25 & $-$60 22 52.1 & 12.65 &  0.30   & B8 III-V     & $-$25\p &  G     & \\
4755-083 & $-$               &  12 55 07.53 & $-$60 21 38.0 & 12.68 &  0.34   & B9 III       & $-$16\p &  G     & \\
4755-084 & IV$-$16           &  12 53 40.60 & $-$60 24 13.0 & 12.75 &  0.25   & B9 III       & $-$15\p &  G     & \\
4755-085 & I$-$34            &  12 53 17.01 & $-$60 20 31.1 & 12.76 &  0.36   & B9 III       & $-$13\p &  G     & \\
4755-086 & $-$               &  12 52 45.06 & $-$60 19 58.5 & 12.79 &  0.39   & A0 III       & $-$35:  &  G     & \\
4755-087 & IV$-$07           &  12 53 33.93 & $-$60 22 41.7 & 12.81 &  0.25   & B8 III       & $-$19\p &  G     & \\
4755-088 & IV$-$56           &  12 53 22.35 & $-$60 27 42.2 & 12.83 &  0.25   & B8 IIIn      & $-$24\p &  G     & \\
4755-089 & $-$               &  12 53 29.99 & $-$60 14 51.1 & 12.84 &  0.36   & B8 III       & $-$19\p &  G     & \\
4755-090 & III$-$09          &  12 53 57.13 & $-$60 22 31.6 & 12.88 &  0.27   & B8 IIIn      & $-$16\p &  G     & \\
4755-091 & II$-$34           &  12 53 48.11 & $-$60 17 45.1 & 12.90 &  0.37   & B8 III       & $-$20\p &  G     & \\
4755-092 & IV$-$35           &  12 53 38.09 & $-$60 25 59.9 & 12.91 &  0.14   & A0 IIp (Si)  & $-$24\p &  G     & \\
4755-093 & III$-$02$^\ast$   &  12 53 44.65 & $-$60 22 32.4 & 12.91 &  0.28   & B8 III-V     & $-$29\p &  G     & \\
4755-094 & IV$-$45           &  12 53 20.30 & $-$60 25 35.1 & 12.93 &  0.31   & B8 III       & $-$30\p &  G     & \\
4755-095 & III$-$33          &  12 53 55.12 & $-$60 26 50.6 & 12.96 &  0.20   & B8 III       & $-$22\p &  G     & \\
4755-096 & II$-$12           &  12 53 50.68 & $-$60 21 22.2 & 12.96 &  0.29   & B8 III       & $-$13\p &  G     & \\
4755-097 & IV$-$21$^\ast$    &  12 53 34.12 & $-$60 23 45.6 & 12.96 &  0.34   & A0 III       & $-$20\p &  G     & \\
4755-098 & IV$-$44           &  12 53 22.02 & $-$60 25 01.2 & 13.00 &  0.25   & B8 IIIn      & $-$20\p &  G     & \\
4755-099 & $-$               &  12 54 28.30 & $-$60 14 45.7 & 13.01 &  0.41   & B8 III       & $-$20\p &  G     & \\
4755-100 & IV$-$01           &  12 53 41.77 & $-$60 22 43.9 & 13.09 &  0.26   & B8 III       & $-$15\p &  G     & \\
4755-101 & $-$               &  12 52 52.93 & $-$60 17 48.8 & 13.10 &  0.28   & B9 II        & $-$22\p &  G     & \\
4755-102 & $-$               &  12 53 53.17 & $-$60 21 24.9 & 13.11 &  0.39   & A2 III       & $-$22\p &  G     & near to II$-$10\\
4755-103 & I$-$28            &  12 52 56.56 & $-$60 20 29.3 & 13.11 &  0.40   & A3 III       & $-$19\p &  G     & \\
4755-104 & III$-$20          &  12 54 03.25 & $-$60 24 01.5 & 13.13 &  0.27   & A0 IIp (Si)  & $-$12:  &  G     & \\
4755-105 & $-$               &  12 54 17.26 & $-$60 17 54.9 & 13.13 &  0.43   & B9 III       & $-$28\p &  G     & \\
4755-106 & $-$               &  12 54 57.47 & $-$60 23 41.4 & 13.16 &  0.32   & B8 III       & $-$18\p &  G     & \\
4755-107 & IV$-$24           &  12 53 23.56 & $-$60 23 12.5 & 13.18 &  0.29   & B8 III-V     & $-$20\p &  G     & \\
4755-108 & I$-$21            &  12 53 09.20 & $-$60 22 03.0 & 13.22 &  0.31   & B8 III-V     & $-$22\p &  G     & \\
\end{longtable}
\end{center}
}

{\scriptsize
\begin{center}
\begin{longtable}{llcccclrll}
\caption[]{NGC6611: Observational Parameters of Target Stars.  
Cross-identifications in column two are from \citet[W, ][]{w61}, 
\citet[K, ][]{k74}, and \citet[T, ][]{t86}.  WFI
photometry is quoted for $V >$ 11.47$^{\rm m}$, brighter than this
there were saturation problems and values are from \citet{hmsm}.
Radial velocities ($v_{\rm r}$) are given in \kms.  Note that stars
6611-001, 6611-005 and 6611-045 are actually outside of the FLAMES
field-of-view.  Photometry for 6611-040 and 6611-045 is also from
\citet{hmsm}.  Instrument codes refer to FEROS (F), Giraffe (G), and ISIS
(I).
\label{6611}} \\
\hline
ID & Alias & $\alpha$(2000) & $\delta$(2000) & $V$ & $B-V$ & Sp. Type & $v_{\rm r}$  & Inst. & Comments \\
\hline 
\endfirsthead
\caption[]{\it{continued}} \\
\hline
ID & Alias & $\alpha$(2000) & $\delta$(2000) & $V$ & $B-V$ & Sp. Type & $v_{\rm r}$  & Inst. & Comments \\
\hline 
\endhead
\hline 
\multicolumn{10}{r}{\it{continued on next page}} \\
\endfoot
\hline 
\multicolumn{10}{l}{Prefixes in the final column refer to: MWC, \citep[Mount Wilson Catalogue, ][]{mwc49}; MCW, \citep{mcw55}; BKP, \citep{bkp}.}\\
\multicolumn{10}{l}{Stars marked with `$\ast$' are those for which {\it definite} visual identifications were not possible from the
\citet{k74} finding chart, with the cross-references taken from \citet{bkp}.}\\
\endlastfoot
\hline
6611-001 & W412           & 18 18 58.71 & $-$13 59 28.1 &\o8.18 &  0.34 & B0 III           &    86\p    & F$+$I      & HD168183, MCW660, BD$-$14$^\circ$4991\\
6611-002 & W205           & 18 18 36.44 & $-$13 48 03.1 &\o8.18 &  0.43 & O4 V((f$^+$))    & 14\p& F$+$I & HD168076, MCW656, BD$-$13$^\circ$4926; H$\alpha$ = wk. P Cyg\\
6611-003 & W197           & 18 18 36.06 & $-$13 47 36.3 &\o8.73 &  0.45 & O6-7 V((f))$+$B0:&    17\p    &  F$+$I     & HD168075, MCW655, BD$-$13$^\circ$4925\\
6611-004 & W401           & 18 18 56.21 & $-$13 48 31.0 &\o8.90 &  0.04 & O8.5 V           &    23\p    &  F$+$I     & HD168137, MCW659, BD$-$13$^\circ$4932\\
6611-005 & $-$            & 18 20 34.10 & $-$13 57 15.8 &\o9.13 &  0.44 & O8 III           &     5\p    &  F$+$I     & HD168504, BD$-$14$^\circ$5005\\
6611-006 & W367           & 18 18 52.70 & $-$13 49 42.6 &\o9.39 &  0.24 & O9.7 IIIp        &  $-$1\p    &  G$+$F$+$I & BD$-$13$^\circ$4930\\
6611-007 & W468           & 18 19 05.58 & $-$13 54 50.3 &\o9.40 &  0.28 & B0.5 V $+$ B1:   &$-$104\p    &  G         & BD$-$13$^\circ$4934\\
6611-008 & W246           & 18 18 40.11 & $-$13 45 18.5 &\o9.46 &  0.82 & O7 II(f)         &    15\p    &  G$+$I     & BD$-$13$^\circ$4927\\
6611-009 & W184           & 18 18 34.52 & $-$13 54 21.9 &\o9.73 &  1.35 & K0 V             & $-$42\p    &  G         & BD$-$13$^\circ$4924\\
6611-010 & W503           & 18 19 11.08 & $-$13 56 42.8 &\o9.75 &  0.49 & B1: e            &     4\p    &  G     & BD$-$13$^\circ$4936, MWC 918; H$\alpha$ = broad em.\\
6611-011 & W314           & 18 18 45.86 & $-$13 46 30.8 &\o9.85 &  0.58 & O9 V             &    17\p    &  G$+$I     & BD$-$13$^\circ$4929\\
6611-012 & W150           & 18 18 29.98 & $-$13 49 57.6 &\o9.85 &  0.48 & B0.5 V           &    14\p    &  G         & BD$-$13$^\circ$4921\\
6611-013 & W125           & 18 18 26.22 & $-$13 50 05.5 & 10.01 &  0.47 & B1 V $+$~?       & $-$92:     &  G         & BD$-$13$^\circ$4920\\
6611-014 & W175           & 18 18 32.75 & $-$13 45 11.9 & 10.09&0.84& O5 V((f$^+$)) $+$late-O& 36\p     &  G$+$I     & BD$-$13$^\circ$4923\\
6611-015 & W280           & 18 18 42.79 & $-$13 46 50.9 & 10.12 &  0.43 & O9.5 Vn          &    23:     &  G$+$I     & BD$-$13$^\circ$4928\\
6611-016 & K599           & 18 19 06.48 & $-$13 57 45.7 & 10.16 &  1.35 & G8 III           & $-$30\p    &  G         & faint companion in WFI image\\
6611-017 & W166           & 18 18 32.24 & $-$13 48 48.0 & 10.37 &  0.57 & O9 V             &    19\p    &  F$+$I     & \\
6611-018 & W002           & 18 18 02.96 & $-$13 44 35.0 & 10.56 &  0.35 & B8 III           &    16\p    &  F         & BD$-$13$^\circ$4914\\
6611-019 & K601           & 18 19 20.01 & $-$13 54 21.3 & 10.68 &  0.36 & B1.5 V           & $-$47\p    &  F         & BD$-$13$^\circ$4937; blended in WFI image\\
6611-020 & W469           & 18 19 04.89 & $-$13 48 20.3 & 10.69 &  0.40 & B0.5 Vn          &    20\p    &  G         & BD$-$13$^\circ$4933\\
6611-021 & W254           & 18 18 40.77 & $-$13 46 52.3 & 10.80 &  0.47 & B1 V             & $-$18\p    &  F         & \\
6611-022 & W235           & 18 18 38.84 & $-$13 46 44.2 & 10.98 &  0.82 & Herbig Be        &   $-$\p    &  G         & MWC 916; H$\alpha$ = broad em.\\
6611-023 & W483           & 18 19 06.51 & $-$13 43 30.3 & 10.99 &  0.41 & B3 V             & $-$19:     &  G         & BD$-$13$^\circ$4935\\
6611-024 & W501           & 18 19 09.95 & $-$13 50 53.0 & 11.19 &  1.67 & K2: V            & $-$13\p    &  G        & \\
6611-025 & W223           & 18 18 37.88 & $-$13 46 35.1 & 11.20 &  0.59 & B1 V             &     8\p    &  G         & \\
6611-026 & W432           & 18 19 00.20 & $-$13 55 34.3 & 11.25 &  0.53 & F0 III           &     0\p    &  G         & \\
6611-027 & W351           & 18 18 50.81 & $-$13 48 12.7 & 11.26 &  0.45 & B1 V             &     0\p    &  F         & \\
6611-028 & W500           & 18 19 09.02 & $-$13 43 14.8 & 11.28 &  0.43 & B5e              &   $-$\p    &  G         & H$\alpha$ = broad em.\\
6611-029 & W161           & 18 18 30.97 & $-$13 43 08.2 & 11.29 &  1.05 & O8.5 V           &  $-$6\p    &  G$+$I    & \\ 
6611-030 & W536           & 18 19 18.49 & $-$13 55 39.8 & 11.46 &  0.22 & B1.5 V $+$~?     &    42\p    &  G         & \\
6611-031 & W407           & 18 18 55.38 & $-$13 39 10.4 & 11.46 &  0.60 & A3 III           &     0\p    &  G         & \\
6611-032 & W239           & 18 18 40.02 & $-$13 54 33.4 & 11.48 &  0.36 & B1.5 V           &     4\p    &  G         & \\
6611-033 & W210           & 18 18 36.99 & $-$13 47 52.7 & 11.50 &  0.54 & B1 V             &     8\p    &  G        & \\
6611-034 & W489           & 18 19 07.33 & $-$13 43 04.6 & 11.52 &  0.54 & B8 III           & $-$18\p    &  G        & \\
6611-035 & W259           & 18 18 40.98 & $-$13 45 29.6 & 11.56 &  0.73 & B0.5 V           &    16\p    &  G        & \\
6611-036 & K592$^\ast$    & 18 19 02.98 & $-$13 36 04.1 & 11.60 &  0.57 & A3 III           &     6\p    &  G        & \\
6611-037 & K600           & 18 19 16.42 & $-$13 54 14.1 & 11.62 &  1.38 & K0 V             &    37\p    &  G        & \\
6611-038 & W520           & 18 19 13.99 & $-$13 52 21.5 & 11.66 &  0.45 & B5 IIIn          & $-$15\p    &  G        & \\
6611-039 & W349           & 18 18 50.59 & $-$13 47 33.9 & 11.69 &  1.35 & G5 III           &     5\p    &  G        & \\
6611-040 & W090           & 18 18 20.20 & $-$13 46 11.9 & 11.73 &  0.38 & B5 V             &    30\p    &  F         & In WFI chip-gap\\
6611-041 & K591           & 18 19 03.23 & $-$13 56 07.3 & 11.75 &  0.44 & B5 V             &     6\p    &  G        & \\
6611-042 & W296           & 18 18 44.68 & $-$13 47 56.3 & 11.81 &  0.49 & B1.5 V           &    14\p    &  G        & \\
6611-043 & W117           & 18 18 24.88 & $-$13 42 42.5 & 11.77 &  1.63 & K0 V             &     5\p    &  G        & \\
6611-044 & W417           & 18 18 57.21 & $-$13 41 37.0 & 11.95 &  1.61 & K0 V             & $-$49\p    &  G        & \\
6611-045 & W584           & 18 18 23.64 & $-$13 36 28.2 & 12.02 &  1.05 & O9 V             &     6\p    &  F         & \\
6611-046 & K583$^\ast$    & 18 18 29.90 & $-$13 36 24.8 & 12.04 &  0.42 & B9 III           &  $-$1\p    &  G        & \\
6611-047 & K611$^\ast$    & 18 19 30.03 & $-$13 44 12.5 & 12.00 &  1.66 & K0 V             & $-$23\p    &  G        & \\
6611-048 & W290           & 18 18 44.86 & $-$13 56 22.2 & 12.12 &  0.46 & B2.5 V           &     8\p    &  G        & \\ 
6611-049 & W455           & 18 19 02.91 & $-$13 47 17.5 & 12.12 &  0.54 & A5 II            & $-$15\p    &  G        & \\
6611-050 & W411           & 18 18 56.97 & $-$13 44 06.5 & 12.08 &  1.63 & K0 V             &    62\p    &  G        & \\
6611-051 & W429           & 18 18 57.38 & $-$13 38 13.3 & 12.12 &  0.81 & F5 V             &    14\p    &  G        & Metal weak?\\
6611-052 & W301           & 18 18 45.00 & $-$13 46 24.8 & 12.19 &  0.58 & B2 V             &     7\p    &  G        & \\
6611-053 & W394           & 18 18 56.21 & $-$13 56 02.0 & 12.19 &  1.18 & G5 III-V         &    13\p    &  G        & \\
6611-054 & K590$^\ast$    & 18 18 50.41 & $-$13 57 04.3 & 12.27 &  0.28 & A0 II            &    48:     &  G        & \\
6611-055 & W440           & 18 19 00.55 & $-$13 46 34.9 & 12.27 &  1.36 & K0 V             &    12\p    &  G        & \\
6611-056 & W484           & 18 19 06.91 & $-$13 45 04.4 & 12.39 &  0.51 & B8 III           & $-$21\p    &  G        & \\
6611-057 & W079           & 18 18 18.38 & $-$13 43 37.2 & 12.44 &  0.63 & A7 II            &     2\p    &  G        & \\
6611-058 & W473           & 18 19 05.73 & $-$13 53 33.4 & 12.46 &  0.24 & A0 II            & $-$10\p    &  G        & \\
6611-059 & W346           & 18 18 49.33 & $-$13 39 24.8 & 12.43 &  1.49 & K0 V             & $-$30\p    &  G        & \\
6611-060 & T636           & 18 19 22.08 & $-$13 40 15.5 & 12.51 &  1.45 & A5 II            &    50:     &  G        & \\
6611-061 & K573$^\ast$    & 18 18 06.99 & $-$13 41 13.3 & 12.54 &  0.77 & F8 V             & $-$13:     &  G        & \\
6611-062 & W289           & 18 18 44.11 & $-$13 48 56.4 & 12.60 &  0.52 & B3 V             &     8\p    &  G        & \\
6611-063 & W444           & 18 19 00.44 & $-$13 42 40.9 & 12.65 &  0.92 & B1.5 V           &     7\p    &  G        & \\
6611-064 & W311           & 18 18 45.60 & $-$13 47 53.1 & 12.78 &  0.55 & B3 V             &    14\p    &  G        & \\
6611-065 & W504           & 18 19 10.32 & $-$13 49 03.7 & 12.80 &  0.40 & B9 III           &    11\p    &  G        & \\
6611-066 & W227           & 18 18 38.42 & $-$13 47 09.0 & 12.83 &  0.62 & B2 V             & $-$50:     &  G        & \\ 
6611-067 & W303           & 18 18 45.88 & $-$13 54 40.1 & 12.85 &  0.65 & F0 III-V         &    30:     &  G        & \\
6611-068 & W472           & 18 19 04.72 & $-$13 44 44.4 & 12.85 &  0.53 & B3 V $+$~?       &     2\p    &  G        & \\
6611-069 & W409           & 18 18 57.39 & $-$13 52 12.1 & 12.89 &  0.41 & B2.5 V           &    17\p    &  G        & \\
6611-070 & W400           & 18 18 55.85 & $-$13 46 54.0 & 12.88 &  0.59 & B9 III           & $-$32:     &  G        & \\
6611-071 & W297           & 18 18 44.54 & $-$13 45 48.1 & 12.89 &  0.67 & B2 Vn            &     1\p    &  G        & \\
6611-072 & W313           & 18 18 46.15 & $-$13 49 23.4 & 12.93 &  0.49 & B5 III           &    23\p    &  G        & \\
6611-073 & W135           & 18 18 27.81 & $-$13 55 04.9 & 12.93 &  0.70 & F0 III-V         &     2\p    &  G        & \\
6611-074 & W272           & 18 18 41.15 & $-$13 37 36.9 & 12.99 &  0.95 & F5 III-V         & $-$46:     &  G        & \\
6611-075 & T639           & 18 19 26.24 & $-$13 47 25.2 & 13.06 &  0.57 & A7 II            &    14\p    &  G        & \\
6611-076 & W163           & 18 18 30.52 & $-$13 37 05.6 & 13.05 &  0.80 & F0 III-V         & $-$34:     &  G        & \\
6611-077 & W174           & 18 18 33.15 & $-$13 51 37.1 & 13.05 &  0.90 & F8 V             & $-$10\p    &  G        & \\
6611-078 & W267           & 18 18 41.71 & $-$13 46 43.8 & 13.11 &  0.52 & B3 V             &     5\p    &  G        & \\
6611-079 & $-$            & 18 19 01.53 & $-$13 35 57.1 & 13.09 &  0.82 & F0 III           &    15\p    &  G        & BKP 29509\\
6611-080 & W222           & 18 18 37.52 & $-$13 43 39.4 & 13.08 &  1.35 & O7 V((f))        &    16\p    &  G        & \\
6611-081 & $-$            & 18 18 42.78 & $-$13 35 57.3 & 13.14 &  0.56 & A7 II            &     2\p    &  G        & BKP 29510\\
6611-082 & W541           & 18 19 19.13 & $-$13 43 52.1 & 13.21 &  0.64 & B1-3 V           &     2\p    &  G        & \\
6611-083 & W036           & 18 18 11.15 & $-$13 45 36.6 & 13.27 &  0.63 & A7 II            &     2\p    &  G        & \\
6611-084 & W345           & 18 18 50.28 & $-$13 53 00.9 & 13.27 &  0.63 & F0 V             &    30\p    &  G        & \\
6611-085 & W336           & 18 18 49.19 & $-$13 48 04.2 & 13.33 &  0.48 & B5 III           &     5\p    &  G        & \\
\end{longtable}
\end{center}
}

\appendix
\section{Accurate stellar cross-identifications}
In the process of compiling published spectral types for stars in
NGC~3293 and NGC~4755, we have discovered several ambiguous
cross-identifications in the literature that we attempt to clarify
here.

\subsection{NGC\,3293-012: HD~92007}
\citet{f3293} listed (his) star \#6 as HD~92007, with the comment that 
both this and star \#26 comprise CPD$-$57$^\circ$3526 in the Cape
Photographic Durchmusterung (CPD).  In the SIMBAD database the two
stars are sensibly referred to as CPD$-$57$^\circ$3526 (Cl* NGC~3293
FEAS 26, in the SIMBAD notation for the Feast targets) and CPD$-$57
3526B (Cl* NGC~3293 FEAS 6), but it is the former star (\#26)
identified as HD~92007, not the latter.  The relevant volume of the
Henry Draper Catalogue \citep{hdcat} in fact lists HD~92007 as
CPD$-$57$^\circ$3527, (\#27 from Feast); this conclusion is supported
from comparisons of our astrometry for Feast \#27 with precessed
positions from the HD Catalogue and we adopt this
cross-identification.  Further confusion is added to the situation by
\citet{ghs77} who give CPD$-$57$^\circ$3540 as HD~92007, instead of
its correct identification of HD~92044 \citep{hdcat}.  Curiously, if
one follows the relevant links from a SIMBAD query for
CPD$-$57$^\circ$3527, one recovers the MK spectral type for HD~92007
from \citet{h56}.

\subsection{NGC\,3293-003: MCW~1181}
Included in their initial study of southern stars as \#1181,
\citet{mcw55} classified the spectrum of CD$-$57$^\circ$3346 as B1~II
(where `CD' is the Cordoba Durchmusterung).
From the quoted spectral type, one can deduce that \citet{f3293}
matched MCW~1181 with his star \#22 (CPD$-$57$^\circ$3506A).  
The SIMBAD database entry for MCW~1181 reveals cross-references to
CD$-$57$^\circ$3346 and, more surprisingly, CPD$-$57$^\circ$3502, a
well-documented M-type supergiant, classified as M0~Iab
\citep[][ \#21]{f3293} and M1.5~Iab-Ib \citep{mk73}.  Despite a
significant literature search, the explicit source of the CPD and CD
cross-references employed by SIMBAD for this star (if one exists) has
eluded us, however we note that both \citet{albers} and \citet{mh84}
referred to the M-supergiant as CD$-$57$^\circ$3346.  Here we limit our
conclusions to the fact that CPD$-$57$^\circ$3502 is not the same
source as MCW~1181 and, assuming that the correct star was observed by
Feast, therefore not CD$-$57$^\circ$3346.

To further complicate matters, when the coordinates from the CD
catalogue (epoch 1875, available from the Centre de Donn${\rm
\acute{e}}$es astronomiques de Strasbourg, CDS) for
CD$-$57$^\circ$3346 are precessed to epoch 1900, they do not tally
with those published by \citet{mcw55} for MCW~1181.  A similar check
for CD$-$57$^\circ$3340 (MCW~1179) finds agreement in the positions,
suggesting possible typographical or numerical errors in the case of
MCW~1181.  

\subsection{NGC\,3293: MCW~1179 \& MCW~1182}
The spectra of MCW~1179 (CD$-$57$^\circ$3340) and MCW~1182
(CD$-$57$^\circ$3348) were both classified by \citet{mcw55} as B0.5~V.
These stars were matched by \citet{f3293} as CPD$-$57$^\circ$3507 (his
\#14) and as CPD$-$57$^\circ$3500 (\#16), which are 3293-019 and 3293-010
respectively in the current study -- although as a result of the
identical classifications it is not clear which CPD star refers to
which MCW target.  Again use of the SIMBAD database highlights
apparent mismatches.  The SIMBAD information for MCW~1179 refers to
Feast \#10, and for MCW~1182 to Feast \#22 (the very star to which
Feast appears to have attributed the spectral type for MCW~1181).  As
for MCW~1181, we were again frustrated in our search for the explicit
source for the cross-references employed by SIMBAD.

\subsection{NGC\,4755-008: I$-$05}
Star I$-$05 from \citet{as58} was identified as CPD$-$59$^\circ$4553 by
\citet{her60}.  However, SIMBAD matches CPD$-$59$^\circ$4553 as star R
from \citet{f4755} and precession of the coordinates in the CPD
catalogue leads to similar conclusions.  \citet{p76} reported $V$ =
8.86 for CPD$-$59$^\circ$4553; a likely explanation is that stars R and
S were spatially unresolved in their study, i.e. the composite object
was the original entry in the CPD catalogue and that the
cross-identification of Hernandez was incorrect.

\bibliographystyle{aa}
\bibliography{2446}

\begin{thebibliography}{98}
\expandafter\ifx\csname natexlab\endcsname\relax\def\natexlab#1{#1}\fi

\bibitem[{{Abel} {et~al.}(2002){Abel}, {Bryan}, \& {Norman}}]{abn02}
{Abel}, T., {Bryan}, G.~L., \& {Norman}, M.~L. 2002, Science, 295, 93

\bibitem[{{Albers}(1972)}]{albers}
{Albers}, H. 1972, ApJ, 176, 623

\bibitem[{{Arp} \& {van Sant}(1958)}]{as58}
{Arp}, H.~C. \& {van Sant}, C.~T. 1958, AJ, 63, 341

\bibitem[{{Bagnulo} {et~al.}(2003){Bagnulo}, {Jehin}, {Ledoux}, {Cabanac},
  {Melo}, \& {Gilmozzi}}]{bcj03}
{Bagnulo}, S., {Jehin}, E., {Ledoux}, C., {et~al.} 2003, ESO Messenger, 114, 10

\bibitem[{{Balona} \& {Crampton}(1974)}]{bc74}
{Balona}, L.~A. \& {Crampton}, D. 1974, MNRAS, 166, 203

\bibitem[{{Becker} {et~al.}(2001){Becker}, {Fan}, {White}, {Strauss},
  {Narayanan}, {Lupton}, {Gunn}, \& {Annis}}]{bfw01}
{Becker}, R.~H., {Fan}, X., {White}, R.~L., {et~al.} 2001, AJ, 122, 2850

\bibitem[{{Belikov} {et~al.}(1999){Belikov}, {Kharchenko}, {Piskunov}, \&
  {Schilbach}}]{bkp}
{Belikov}, A.~N., {Kharchenko}, N.~V., {Piskunov}, A.~E., \& {Schilbach}, E.
  1999, A\&AS, 134, 525

\bibitem[{{Bidelman}(1954)}]{b54}
{Bidelman}, W.~P. 1954, PASP, 66, 249

\bibitem[{{Blaauw}(1956)}]{b56}
{Blaauw}, A. 1956, ApJ, 123, 408

\bibitem[{{Blaauw} \& {Morgan}(1953)}]{bm53}
{Blaauw}, A. \& {Morgan}, W.~W. 1953, ApJ, 117, 256

\bibitem[{{Blecha} {et~al.}(2003){Blecha}, {North}, {Royer}, \&
  {Simond}}]{girbldrs}
{Blecha}, A., {North}, P., {Royer}, F., \& {Simond}, G. 2003, BLDR Software -
  Reference Manual, 1st edn.

\bibitem[{{Bloom} {et~al.}(2002){Bloom}, {Kulkarni}, {Price}, {Reichart},
  {Galama}, {Schmidt}, \& {Frail}}]{bk02}
{Bloom}, J.~S., {Kulkarni}, S.~R., {Price}, P.~A., {et~al.} 2002, ApJ, 572, 45L

\bibitem[{{Bosch} {et~al.}(1999){Bosch}, {Morrell}, \& {Niemela}}]{bmn99}
{Bosch}, G.~L., {Morrell}, N.~I., \& {Niemela}, V.~S. 1999, RevMex, 35, 85

\bibitem[{{Bouret} {et~al.}(2003){Bouret}, {Lanz}, {Hillier}, {Heap}, {Hubeny},
  {Lennon}, {Smith}, \& {Evans}}]{jc03}
{Bouret}, J.-C., {Lanz}, T., {Hillier}, D.~J., {et~al.} 2003, ApJ, 595, 1182

\bibitem[{{Bromm} {et~al.}(2002){Bromm}, {Coppi}, \& {Larson}}]{bcl02}
{Bromm}, V., {Coppi}, P.~S., \& {Larson}, R.~B. 2002, ApJ, 564, 23

\bibitem[{{Cannon} \& {Pickering}(1919)}]{hdcat}
{Cannon}, A.~J. \& {Pickering}, E.~C. 1919, Annals Harvard Obs., 94, 1

\bibitem[{{Conti} \& {Leep}(1974)}]{cl74}
{Conti}, P.~S. \& {Leep}, E.~M. 1974, ApJ, 193, 113

\bibitem[{{Crowther} {et~al.}(2002){Crowther}, {Hillier}, {Evans}, {Fullerton},
  {De Marco}, \& {Willis}}]{paul}
{Crowther}, P.~A., {Hillier}, D.~J., {Evans}, C.~J., {et~al.} 2002, ApJ, 579,
  774

\bibitem[{{Dachs} \& {Kaiser}(1984)}]{dk84}
{Dachs}, J. \& {Kaiser}, D. 1984, A\&AS, 58, 411

\bibitem[{{de Waard} {et~al.}(1984){de Waard}, {van Genderen}, \&
  {Bijleveld}}]{dewaard}
{de Waard}, G.~J., {van Genderen}, A.~M., \& {Bijleveld}, W. 1984, A\&AS, 56,
  373

\bibitem[{{de Winter} {et~al.}(1997){de Winter}, {Koulis}, {Th$\acute{\rm e}$},
  {van den Ancker}, {P$\acute{\rm e}$rez}, \& {Bibo}}]{dW97}
{de Winter}, D., {Koulis}, C., {Th$\acute{\rm e}$}, P.~S., {et~al.} 1997,
  A\&AS, 121, 223

\bibitem[{{Duch$\hat{\rm e}$ne} {et~al.}(2001){Duch$\hat{\rm e}$ne}, {Simon},
  {Eisl$\ddot{\rm o}$ffel}, \& {Bouvier}}]{ds01}
{Duch$\hat{\rm e}$ne}, G., {Simon}, T., {Eisl$\ddot{\rm o}$ffel}, J., \&
  {Bouvier}, J. 2001, A\&A, 379, 147

\bibitem[{{Evans} {et~al.}(2004{\natexlab{a}}){Evans}, {Crowther}, {Fullerton},
  \& {Hillier}}]{ecfh}
{Evans}, C.~J., {Crowther}, P.~A., {Fullerton}, A.~W., \& {Hillier}, D.~J.
  2004{\natexlab{a}}, ApJ, 610, 1021

\bibitem[{{Evans} \& {Howarth}(2003)}]{eh03}
{Evans}, C.~J. \& {Howarth}, I.~D. 2003, MNRAS, 345, 1223

\bibitem[{{Evans} {et~al.}(2004{\natexlab{b}}){Evans}, {Howarth}, {Irwin},
  {Burnley}, \& {Harries}}]{eh04}
{Evans}, C.~J., {Howarth}, I.~D., {Irwin}, M.~J., {Burnley}, A.~W., \&
  {Harries}, T.~J. 2004{\natexlab{b}}, MNRAS, 353, 601

\bibitem[{{Evans} {et~al.}(2004{\natexlab{c}}){Evans}, {Lennon}, {Walborn},
  {Trundle}, \& {Rix}}]{elw04}
{Evans}, C.~J., {Lennon}, D.~J., {Walborn}, N.~R., {Trundle}, C., \& {Rix},
  S.~A. 2004{\natexlab{c}}, PASP, 116, 909

\bibitem[{{Feast}(1958)}]{f3293}
{Feast}, M.~W. 1958, MNRAS, 118, 618

\bibitem[{{Feast}(1963)}]{f4755}
{Feast}, M.~W. 1963, MNRAS, 126, 11

\bibitem[{{Feinstein} \& {Marraco}(1980)}]{fm90}
{Feinstein}, A. \& {Marraco}, H.~G. 1980, PASP, 92, 266

\bibitem[{{Galama} {et~al.}(1998){Galama}, {Vreeswijk}, {van Paradijs},
  {Kouveliotou}, {Augusteijn}, \& {Bohnhardt}}]{gal98}
{Galama}, T.~J., {Vreeswijk}, P.~M., {van Paradijs}, J., {et~al.} 1998, Nature,
  395, 670

\bibitem[{{Garrison} {et~al.}(1977){Garrison}, {Hiltner}, \& {Schild}}]{ghs77}
{Garrison}, R.~F., {Hiltner}, W.~A., \& {Schild}, R.~E. 1977, ApJS, 35, 111

\bibitem[{{Haehnelt} {et~al.}(2001){Haehnelt}, {Madau}, {Kudritzki}, \&
  {Haardt}}]{hmkh01}
{Haehnelt}, M.~G., {Madau}, P., {Kudritzki}, R.-P., \& {Haardt}, F. 2001, ApJ,
  549, 151L

\bibitem[{{Heger} \& {Langer}(2000)}]{hl00}
{Heger}, A. \& {Langer}, N. 2000, ApJ, 544, 1016

\bibitem[{{Herbig}(1960)}]{h60}
{Herbig}, G.~H. 1960, ApJS, 4, 337

\bibitem[{{Herbig} \& {Dahm}(2001)}]{hd01}
{Herbig}, G.~H. \& {Dahm}, S.~E. 2001, PASP, 113, 195

\bibitem[{{Herbst} \& {Miller}(1982)}]{hm82}
{Herbst}, W. \& {Miller}, D.~P. 1982, AJ, 87, 1478

\bibitem[{{Hern${\rm \acute{a}}$ndez}(1960)}]{her60}
{Hern${\rm \acute{a}}$ndez}, C. 1960, PASP, 72, 416

\bibitem[{{Hill} {et~al.}(1986){Hill}, {Walker}, \& {Yang}}]{hwy86}
{Hill}, G.~M., {Walker}, G. A.~H., \& {Yang}, S. 1986, PASP, 98, 1186

\bibitem[{{Hillenbrand} {et~al.}(1993){Hillenbrand}, {Massey}, {Strom}, \&
  {Merrill}}]{hmsm}
{Hillenbrand}, L.~A., {Massey}, P., {Strom}, S.~E., \& {Merrill}, K.~M. 1993,
  AJ, 106, 1906

\bibitem[{{Hillier} {et~al.}(2003){Hillier}, {Lanz}, {Heap}, {Hubeny}, {Smith},
  {Evans}, {Lennon}, \& {Bouret}}]{jdh03}
{Hillier}, D.~J., {Lanz}, T., {Heap}, S.~R., {et~al.} 2003, ApJ, 588, 1039

\bibitem[{{Hiltner} {et~al.}(1969){Hiltner}, {Garrison}, \& {Schild}}]{hgs69}
{Hiltner}, W.~A., {Garrison}, R.~F., \& {Schild}, R.~E. 1969, ApJ, 157, 313

\bibitem[{{Hiltner} \& {Morgan}(1969)}]{hm69}
{Hiltner}, W.~A. \& {Morgan}, W.~W. 1969, AJ, 74, 1152

\bibitem[{{Hjorth} {et~al.}(2003){Hjorth}, {Sollerman}, {Moller}, {Fynbo},
  {Woosley}, {Kouveliotou}, \& {Tanvir}}]{hjorth}
{Hjorth}, J., {Sollerman}, J., {Moller}, P., {et~al.} 2003, Nature, 423, 847

\bibitem[{{Hoffleit}(1956)}]{h56}
{Hoffleit}, D. 1956, ApJ, 124, 61

\bibitem[{{Irwin} \& {Lewis}(2001)}]{mji_wfc}
{Irwin}, M.~J. \& {Lewis}, J. 2001, NewAR, 45, 105

\bibitem[{{Kamp}(1974)}]{k74}
{Kamp}, L.~W. 1974, A\&AS, 16, 1

\bibitem[{{Kaufer} {et~al.}(1999){Kaufer}, {Stahl}, \& {Tubbesing}}]{ak99}
{Kaufer}, A., {Stahl}, O., \& {Tubbesing}, S. e.~a. 1999, The ESO Messenger,
  95, 8

\bibitem[{{Kudritzki}(2002)}]{kud02}
{Kudritzki}, R.-P. 2002, ApJ, 577, 389

\bibitem[{{Kudritzki} {et~al.}(1995){Kudritzki}, {Lennon}, \& {Puls}}]{klp95}
{Kudritzki}, R.-P., {Lennon}, D.~J., \& {Puls}, J. 1995, in Science with the
  VLT, ed. J.~R. {Walsh} \& I.~J. {Danziger} (Springer, Berlin), 246

\bibitem[{{Kudritzki} {et~al.}(1987){Kudritzki}, {Pauldrach}, \& {Puls}}]{kpp}
{Kudritzki}, R.-P., {Pauldrach}, A. W.~A., \& {Puls}, J. 1987, A\&A, 173, 293

\bibitem[{{Kudritzki} {et~al.}(1999){Kudritzki}, {Puls}, {Lennon}, {Venn},
  {Reetz}, {Najarro}, {McCarthy}, \& {Herrero}}]{k99}
{Kudritzki}, R.-P., {Puls}, J., {Lennon}, D.~J., {et~al.} 1999, A\&A, 350, 970

\bibitem[{{Lennon} {et~al.}(1992){Lennon}, {Dufton}, \& {Fitzsimmons}}]{ldf92}
{Lennon}, D.~J., {Dufton}, P.~L., \& {Fitzsimmons}, A. 1992, A\&AS, 94, 569

\bibitem[{{Lennon} {et~al.}(1993){Lennon}, {Mazzali}, {Pasian}, {Bonifacio}, \&
  {Castellani}}]{l93}
{Lennon}, D.~J., {Mazzali}, P.~A., {Pasian}, P., {Bonifacio}, P., \&
  {Castellani}, V. 1993, SSRv, 66, 169

\bibitem[{{Lesh}(1968)}]{l68}
{Lesh}, J.~R. 1968, ApJS, 17, 371

\bibitem[{{Maeder} \& {Meynet}(2000)}]{mm00}
{Maeder}, A. \& {Meynet}, G. 2000, A\&A, 361, 101

\bibitem[{{Maeder} \& {Meynet}(2001)}]{mm01}
{Maeder}, A. \& {Meynet}, G. 2001, A\&A, 373, 555

\bibitem[{{Markova} {et~al.}(2004){Markova}, {Puls}, {Repolust}, \&
  {Markov}}]{mp04}
{Markova}, N., {Puls}, J., {Repolust}, T., \& {Markov}, H. 2004, A\&A, 413, 693

\bibitem[{{McGregor} \& {Hyland}(1984)}]{mh84}
{McGregor}, P.~J. \& {Hyland}, A.~R. 1984, ApJ, 277, 149

\bibitem[{{Merrill} \& {Burwell}(1949)}]{mwc49}
{Merrill}, P.~W. \& {Burwell}, C.~G. 1949, ApJ, 110, 387

\bibitem[{{Momany} {et~al.}(2001){Momany}, {Vandame}, {Zaggia}, {Mignani}, {da
  Costa}, {Arnouts}, {Groenewegen}, {Hatziminaoglou}, {Madejsky},
  {Rit$\acute{\rm e}$}, {Schirmer}, \& {Slijkhuis}}]{eis}
{Momany}, Y., {Vandame}, B., {Zaggia}, S., {et~al.} 2001, A\&A, 379, 436

\bibitem[{{Morgan}(1933)}]{m33}
{Morgan}, W.~W. 1933, ApJ, 77, 330

\bibitem[{{Morgan} {et~al.}(1955){Morgan}, {Code}, \& {Whitford}}]{mcw55}
{Morgan}, W.~W., {Code}, A.~D., \& {Whitford}, A.~E. 1955, ApJS, 2, 41

\bibitem[{{Morgan} \& {Keenan}(1973)}]{mk73}
{Morgan}, W.~W. \& {Keenan}, P.~C. 1973, ARA\&A, 11, 29

\bibitem[{{Morgan} {et~al.}(1943){Morgan}, {Keenan}, \& {Kellman}}]{mkk}
{Morgan}, W.~W., {Keenan}, P.~C., \& {Kellman}, E. 1943, An atlas of stellar
  spectra (Chicago Univ. Press)

\bibitem[{{Morgan} {et~al.}(1953){Morgan}, {Whitford}, \& {Code}}]{mwc53}
{Morgan}, W.~W., {Whitford}, A.~E., \& {Code}, A.~D. 1953, ApJ, 118, 318

\bibitem[{{Oudmaijer} {et~al.}(1997){Oudmaijer}, {Drew}, {Barlow}, {Crawford},
  \& {Proga}}]{o97}
{Oudmaijer}, R.~D., {Drew}, J.~E., {Barlow}, M.~J., {Crawford}, I.~A., \&
  {Proga}, D. 1997, MNRAS, 291, 110

\bibitem[{{Parker} {et~al.}(1992){Parker}, {Garmany}, {Massey}, \&
  {Walborn}}]{p92}
{Parker}, J.~W., {Garmany}, C.~D., {Massey}, P., \& {Walborn}, N.~R. 1992, AJ,
  103, 1205

\bibitem[{{Perry} {et~al.}(1976){Perry}, {Franklin}, {Landolt}, \&
  {Crawford}}]{p76}
{Perry}, C.~L., {Franklin}, C.~B., {Landolt}, A.~U., \& {Crawford}, D.~L. 1976,
  AJ, 81, 632

\bibitem[{{Petrie}(1965)}]{p65}
{Petrie}, R.~M. 1965, Publ. Dom. Astrophys. Obs. Victoria, 12, 317

\bibitem[{{Pettini} {et~al.}(2002){Pettini}, {Ellison}, {Bergeron}, \&
  {Petitjean}}]{max}
{Pettini}, M., {Ellison}, S.~L., {Bergeron}, J., \& {Petitjean}, P. 2002, A\&A,
  391, 21

\bibitem[{{Puls} {et~al.}(1996){Puls}, {Kudritzki}, {Herrero}, {Pauldrach},
  {Haser}, {Lennon}, {Gabler}, {Voels}, {Vilchez}, {Wachter}, \&
  {Feldmeier}}]{p96}
{Puls}, J., {Kudritzki}, R.-P., {Herrero}, A., {et~al.} 1996, A\&A, 305, 171

\bibitem[{{Repolust} {et~al.}(2004){Repolust}, {Puls}, \& {Herrero}}]{rp04}
{Repolust}, T., {Puls}, J., \& {Herrero}, A. 2004, A\&A, 415, 349

\bibitem[{{Rix} {et~al.}(2004){Rix}, {Pettini}, {Leitherer}, {Bresolin},
  {Kudritzki}, \& {Steidel}}]{rix}
{Rix}, S., {Pettini}, M., {Leitherer}, C., {et~al.} 2004, ApJ, 615, 98

\bibitem[{{Roman} \& {Morgan}(1950)}]{rm50}
{Roman}, N.~G. \& {Morgan}, W.~W. 1950, ApJ, 111, 426

\bibitem[{{Sanner} {et~al.}(2001){Sanner}, {Brunzendorf}, {Will}, \&
  {Geffert}}]{sb01}
{Sanner}, J., {Brunzendorf}, J., {Will}, J.-M., \& {Geffert}, M. 2001, A\&A,
  369, 511

\bibitem[{{Schild}(1970)}]{s70}
{Schild}, R.~E. 1970, ApJ, 161, 855

\bibitem[{{Shapley} {et~al.}(2004){Shapley}, {Erb}, {Pettini}, {Steidel}, \&
  {Adelberger}}]{shap04}
{Shapley}, A.~E., {Erb}, D.~K., {Pettini}, M., {Steidel}, C.~C., \&
  {Adelberger}, K.~L. 2004, ApJ, 612, 108

\bibitem[{{Smartt} {et~al.}(2004){Smartt}, {Maund}, {Hendry}, {Tout},
  {Gilmore}, {Mattila}, \& {Benn}}]{smh04}
{Smartt}, S.~J., {Maund}, J.~R., {Hendry}, M.~A., {et~al.} 2004, Science, 303,
  499

\bibitem[{{Th$\acute{\rm e}$} {et~al.}(1990){Th$\acute{\rm e}$}, {de Winter},
  {Feinstein}, \& {Westerlund}}]{the90}
{Th$\acute{\rm e}$}, P.~S., {de Winter}, D., {Feinstein}, A., \& {Westerlund},
  B.~E. 1990, A\&AS, 82, 319

\bibitem[{{Trundle} \& {Lennon}(2005)}]{tl05}
{Trundle}, C. \& {Lennon}, D.~J. 2005, A\&A, astro-ph/0501228

\bibitem[{{Trundle} {et~al.}(2004){Trundle}, {Lennon}, {Puls}, \&
  {Dufton}}]{tl04}
{Trundle}, C., {Lennon}, D.~J., {Puls}, J., \& {Dufton}, P.~L. 2004, A\&A, 417,
  217

\bibitem[{{Tucholke} {et~al.}(1986){Tucholke}, {Geffert}, \& {Th$\acute{\rm
  e}$}}]{t86}
{Tucholke}, H.-J., {Geffert}, M., \& {Th$\acute{\rm e}$}, P.~S. 1986, A\&AS,
  66, 311

\bibitem[{{Turner} {et~al.}(1980){Turner}, {Grieve}, {Herbst}, \&
  {Harris}}]{turn80}
{Turner}, D.~G., {Grieve}, G.~R., {Herbst}, W., \& {Harris}, W.~E. 1980, AJ,
  85, 1193

\bibitem[{{V$\acute{\rm a}$zquez} {et~al.}(2004){V$\acute{\rm a}$zquez},
  {Leitherer}, {Heckman}, {Lennon}, {de Mello}, {Meurer}, \& {Martin}}]{vlh04}
{V$\acute{\rm a}$zquez}, G.~A., {Leitherer}, C., {Heckman}, T.~M., {et~al.}
  2004, ApJ, 600, 162

\bibitem[{{Vink} {et~al.}(2001){Vink}, {de Koter}, \& {Lamers}}]{v01}
{Vink}, J.~S., {de Koter}, A., \& {Lamers}, H. J. G. L.~M. 2001, A\&A, 369, 574

\bibitem[{{Walborn}(1972)}]{w72}
{Walborn}, N.~R. 1972, AJ, 77, 312

\bibitem[{{Walborn}(1973)}]{w73}
{Walborn}, N.~R. 1973, AJ, 78, 1067

\bibitem[{{Walborn}(1976)}]{w76}
{Walborn}, N.~R. 1976, ApJ, 205, 419

\bibitem[{{Walborn}(1982)}]{w82}
{Walborn}, N.~R. 1982, AJ, 87, 1300

\bibitem[{{Walborn}(1983)}]{w83}
{Walborn}, N.~R. 1983, ApJ, 268, 195

\bibitem[{{Walborn} \& {Fitzpatrick}(1990)}]{wf90}
{Walborn}, N.~R. \& {Fitzpatrick}, E.~L. 1990, PASP, 102, 379

\bibitem[{{Walborn} {et~al.}(2000){Walborn}, {Lennon}, {Heap}, {Lindler},
  {Smith}, {Evans}, \& {Parker}}]{wal00}
{Walborn}, N.~R., {Lennon}, D.~J., {Heap}, S.~R., {et~al.} 2000, PASP, 112,
  1243

\bibitem[{{Walborn} {et~al.}(2004){Walborn}, {Morrell}, {Howarth}, {Crowther},
  {Lennon}, {Massey}, \& {Arias}}]{wal04}
{Walborn}, N.~R., {Morrell}, N.~I., {Howarth}, I.~D., {et~al.} 2004, ApJ, 608,
  1028

\bibitem[{{Walker}(1961)}]{w61}
{Walker}, M.~F. 1961, ApJ, 133, 438

\bibitem[{{Welsh}(1984)}]{welsh84}
{Welsh}, B.~Y. 1984, MNRAS, 207, 167

\bibitem[{{Wilson}(1937)}]{w37}
{Wilson}, O.~C. 1937, PASP, 49, 338

\bibitem[{{Wyithe} \& {Loeb}(2003)}]{wl03}
{Wyithe}, J. S.~B. \& {Loeb}, A. 2003, ApJ, 586, 693

\bibitem[{{Yadav} \& {Sagar}(2001)}]{ys01}
{Yadav}, R. K.~S. \& {Sagar}, R. 2001, MNRAS, 328, 370

\end{thebibliography}

\end{document}